\documentclass{article}

\usepackage{arxiv}

\usepackage[utf8]{inputenc} 
\usepackage[T1]{fontenc}    
\usepackage{hyperref}       
\usepackage{url}            
\usepackage{booktabs}       
\usepackage{amsfonts,amsmath,amssymb}       
\usepackage{nicefrac}       
\usepackage{microtype}      
\usepackage{lipsum}
\usepackage{graphicx,tabularx}
\usepackage{physics}
\usepackage{epstopdf, epsfig}
\usepackage{newtxtext}
\usepackage{newtxmath}
\usepackage{natbib}
\usepackage{soul}
\usepackage{ragged2e} 
\usepackage{multirow}
\usepackage[dvipsnames]{xcolor}
\usepackage[labelformat=simple]{subcaption}
\usepackage{caption}
\usepackage{booktabs}
\usepackage[skip=1pt plus1pt, indent=10pt]{parskip}
 
\hypersetup{
    colorlinks = true,
    urlcolor   = blue,
    citecolor  = black,
}

\newcommand{\RomanNumeralCaps}[1]

\title{Boundary layer flow dynamics of propulsive flapping foils with increasing Reynolds numbers}

\author{
 Andhini N Zurman-Nasution\\
  School of Engineering\\
  University of Southampton\\
  Southampton SO16 7QF, UK\\
  \texttt{A.Zurman-Nasution@soton.ac.uk} \\
   \And
 Gabriel D. Weymouth \\
  Ship Hydromechanics and Structures\\
  TU Delft\\
  The Netherlands\\
  \texttt{G.D.Weymouth@tudelft.nl} \\
  \And
 Bharath Ganapathisubramani \\
  University of Southampton\\
  Southampton SO16 7QF, UK\\
  \texttt{G.Bharath@soton.ac.uk} \\
}

\begin{document}
\maketitle
\begin{abstract}
We study flapping foils at optimally propulsive Strouhal number $St=0.3$ with increasing chord-based Reynolds number at $Re = 10^4$, $10^5$, and $10^6$ to examine changes in their unsteady boundary layers. Despite being prescribed the same freestream, the inner boundary layer characteristics exhibit different trends due to the generation of leading-edge vortices (LEVs) and their advection into the downstream flow. Propulsive flapping foils show an extended laminar region known as the \textit{relaminarization}, during which the velocity profiles deviate from the standard log law. This \textit{relaminarization} is accompanied by a significant decrease in the cyclic fluctuation of the wall friction coefficient and an increase in the shape factor while the freestream velocity increases under favourable pressure gradient conditions. This phenomenon intensifies with increasing $Re$. We found that higher $Re$ produces smaller LEVs in greater quantities, with a more rapid but stable breakdown, without resulting in a more chaotic turbulent downstream. This study strongly indicates that the \textit{relaminarization} could extend beyond $Re=10^6$ as predicted by \citet{Fukagata2023}. The results support the potential for further exploration of flapping foils at high $Re$ for noise and drag reduction. 
\end{abstract}


\section{introduction}

The interest in flapping foil studies has been increasing for more than two decades in the fields of optimal propulsion for aerial and underwater robots, swimming/flying gaits, and recently for drag/noise reduction and flow control. To apply this research at an industrial scale will require a massive increase in the size of the foil and the Reynolds number of the flow. However, high Reynolds number flapping foil flow has not been extensively studied, either numerically or experimentally, motivating an in-depth investigation. 

Flapping foils are reported to be optimally propulsive in the range of $0.2<St<0.4$ \cite{Eloy2012,Triantafyllou1993} for swimming animals with oscillating lift-based propulsion. These $St$ ranges generate the highest efficiency and exhibit two-dimensional (2-D) flow with relatively small Leading-Edge Vortices (LEVs) for pitching-heaving kinematics \citep{Zurman2020}. The flow surrounding a flapping foil is affected by its motion and studying its effect on the boundary layer region foil is expected to reveal the details comprehensively. 

Many studies have been published for the characteristics of the boundary layer flow that we can relate to flapping foil flow, such as the flow over a flat plate and channel flow \citep{Clauser1954}, curved flow and stationary foils  \citep{Coles1956,Mukund2006,Devenport2022,Coles1956}, unsteady flow and boundary layer separation \citep{Ambrogi2022,Na1998}. These studies describe the changes in the flow due to pressure gradients, mostly at the constant zero, adverse, and favourable pressure gradients, abbreviated as ZPG, APG, and FPG consecutively. As the name, the ZPG indicates no pressure change, or $\dv*{Cp}{x}=0$ where $Cp$ is the pressure coefficient. Meanwhile, the APG indicates the increase of pressure nominal to the flow direction ($\dv*{Cp}{x}>0$) and the opposite for FPG  ($\dv*{Cp}{x}<0$). APG also indicates the local flow deceleration, while FPG is related to the acceleration.

As the flow over a foil surface is curved, we are interested in several studies about turbulent flow under the strong influence of pressure gradient. Several studies by \citet{Patel1968, Warnack1998, Mukund2006} emphasized the phenomenon of \textit{relaminarization} under the influence of a strong favourable pressure gradient. \cite{Narasimha1978} called this phenomenon the \textit{reversion} as the result of pressure forces domination over slowly responding Reynolds stresses in the outer region accompanied by laminar generation inside the boundary layer stabilized by the flow acceleration. The characteristics of flow undergoing FPG are thus related to the stabilization by acceleration inside the boundary layer, or vice versa, the APG is known as the flow destabilizing. The variables to indicate this \textit{relaminarization} phenomenon is $\Delta_P = -\dv*{p}{x}^+ = -\frac{\nu}{\rho u_{\tau}^3} \frac{dp}{dx}$ that has characteristics of inner layer as it involves $u_\tau$ \citep{Uzun2021} values at $\Delta_P>0.018$ \citep{Patel1968}. Others use an acceleration parameter such as \cite{Mukund2006} $K=\frac{\nu}{\rho u_e^2} \frac{u_e}{dx} $ that is not universal as it involves the parameter of boundary-layer edge velocity $u_e$ affected by the freestream.

An essential feature of flapping foil flow is the generation of the Leading-Edge Vortices (LEVs) that resemble the separation bubbles. There are more studies using experiments involving turbulent separation bubbles that we can learn about the effect of pressure gradient changes on turbulent flow, such as the blow-and-suction experiments in channel flow \citep{Na1998}. Recently \citet{ Ambrogi2022} concludes that the flow reversal leads to a hysteresis effect at a certain reduced frequency. There is also a review study reported on multiple blow-and-suction experiments by \citet{Fukagata2023} for drag reduction. At a generally low Reynolds number flow, the LEV generated by propulsive flapping foil is similar to a laminar separation bubble or a 2-D LEV. However, in a higher Reynolds number case or large-scale flapping foil, the LEV is expected to behave like a turbulent separation bubble with 3-D flow characteristics.

Only a limited number of studies have focused on the boundary layer of a flapping foil or a swimming animal, mostly because of the setup cost and extensive requirements to zoom in on the region near the wall boundary, both in experiments and numerical simulations. The higher the Reynolds number, the more challenging and complex this study can become. Large mammals, such as dolphins and whales swim at a range of body length $Re \approx 10^6$ and beyond, indicating that turbulent flow most probably occurs on their flapping flukes. Research on smaller Reynolds number ranges has been conducted through experiments such as the fish-robot experiment by \cite{Triantafyllou2002}, and even using live fishes \citet{Yanase2015,Taneda1974, Anderson2001}. We address this gap by extensively studying the boundary layer of a propulsive flapping foil, focusing on the changes in inner boundary layer parameters affected by the outer layer forcing or the freestream, and how they correlate with increasing $Re$.

\section{Methodology}
\subsection{Simulation arrangement for flapping foils}
A NACA0016 with chord length $C$ and foil thickness $0.16C$ oscillates sinusoidally with combined pitch and heave motions. The pitch pivot point is at $0.25C$ from the foil leading edge. The foil is placed in a uniform flow with inflow speed $U_{\infty}$ and fluid density $\rho$. The \textit{ chord-length based} Reynolds number ($Re =U_{\infty}C/\nu$) is progressively increased from $10^4$ to $10^5$ to $10^6$, with $\nu$ as kinematic viscosity. Other bases than the chord length will be provided as a lowercase index of $Re$. 
We study the optimal propulsive Strouhal number $St=2Af/U_{\infty}=0.3$ where $2A$ is the peak-to-peak amplitude at the trailing edge and $f$ is the flapping frequency. This $St$ is within the range where the least spanwise perturbation and the flow exhibit two-dimensionality at low $Re$ \citep{Zurman2020}.

The foil oscillates with sinusoidal kinematics constructed from heaving $H(t)$ and pitching $\theta(t)$ motions as
\begin{align}
	H(t) &= A \sin(2 \pi f t)\label{heaving} \\
	\theta(t) &= \theta_{0} \sin(2 \pi f t + \pi/2) \label{pitching} \\
	\theta_{0} &= \sin^{-1}\big(A/(0.75C)\big) \label{pitching_amp}
\end{align}
where $A=0.25C$. We fixed the flapping reduced frequency $k=fC/U_{\infty}=0.6$. This relatively high reduced frequency allowed us to minimize the peak-to-peak amplitude for the trade-off of our computing capability for the simulation at $Re=10^6$ while keeping maximum efficiency $\eta$ and thrust coefficient $C_T$ at the chosen $St$. The hydrodynamics performances are tabulated in Tab.\ref{tab:stats}). The effective angle of attack $\alpha(t)=-(\tan^{-1}\dv{H(t)}{t}-\theta(t))$ defines the evolution of foil's angle of attack along the cycle \citep{Hover2004}, where the maximum $\alpha_{eff}=\max(|\alpha(t)|)=23.83^{\circ}$ for this study.

\begin{table}
  \begin{center}
  \begin{tabular}{cccc}
    \toprule
      $Re$ 	&  Mean thrust Coefficients ($\overline{C_T}$) & Mean power Coefficients ($\overline{C_{Pow}}$) & Efficiency $\eta$ \\
      $10^6$  & 0.61 & 1.52 & 0.40\\
      $10^5$ 	& 0.65 & 1.66 & 0.39\\
	 $10^4$  & 0.60 & 1.60 &  0.38\\
    \bottomrule
  \end{tabular}
  \caption{Hydrodynamic performance statistics at $St=0.3$. $C_T=-\frac{F_X}{0.5 \rho S_p U_{\infty}^2}$,  $C_{Pow}= \frac{P}{0.5 \rho S_p U_{\infty}^3}$ and  $\eta=\frac{\overline{C_T}}{\overline{C_{Pow}}}$, where $F_X$ is integrated streamwise force, $S_p$ foil's planform area, $P$ integrated power and $\rho $ density. The overline symbol is the total cycle average. $C_T$ includes both pressure and viscous forces.} 
  \label{tab:stats}
  \end{center}
\end{table}

\subsection{Solver and simulation convergence}

\begin{figure}
	\centering
	\begin{subfigure}[b]{.5\textwidth}
    	\includegraphics[width=1\textwidth]{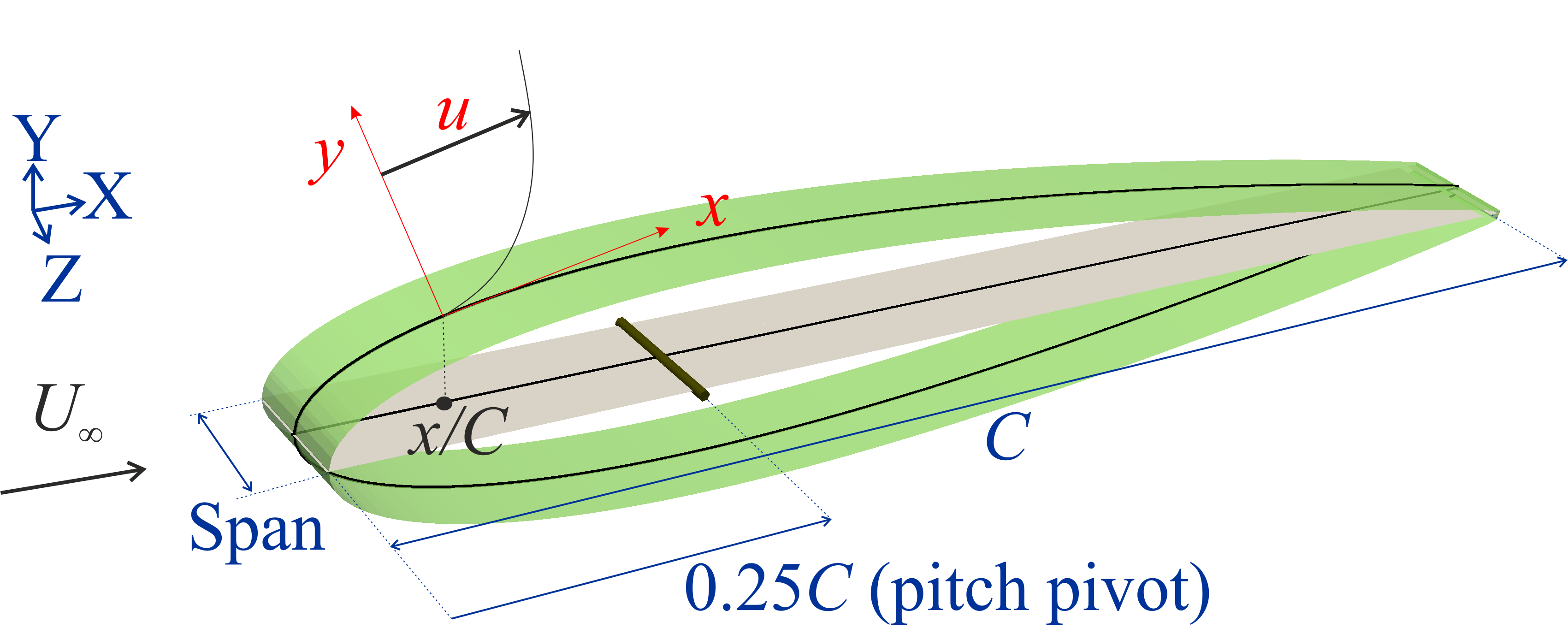}
        \subcaption{}
	\end{subfigure}
	\begin{subfigure}[b]{.75\textwidth}
   	 \includegraphics[width=1\textwidth]{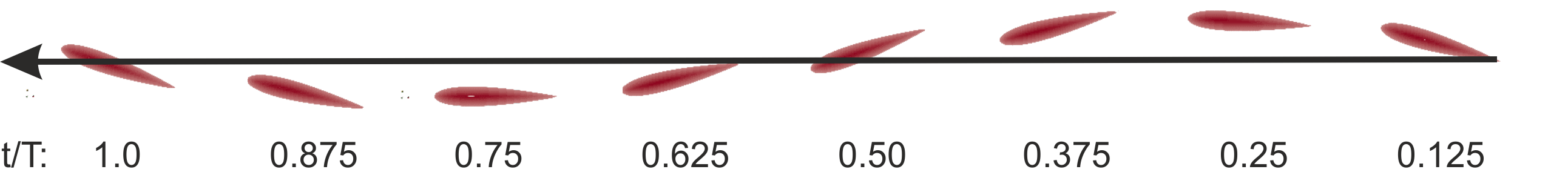}
        \subcaption{}
	\end{subfigure}
	\caption{(a) Representative of foil and the streamwise velocity $U$ with its normal ($y$) and tangential ($x$) axes. (a) Flapping foil phases in time $t$ over a cycle period $T$. For discussion, phases $t$ correspond to the upper side of the foil, while the lower part experiences $t \pm 0.5T$.  }
	\label{fig:phase}
 \end{figure}  

This study uses Boundary Data Immersion Method (BDIM) solver package with second-order accuracy in space \cite{Maertens2014} improved from \cite{Weymouth2011} near the solid-fluid boundary. BDIM uses implicit Large Eddy Simulation (iLES), Cartesian grids, finite volume, and Quick scheme that has been proven to accurately predict the integrated forces on moving bodies in \cite{Maertens2014} up to $Re=10^5$. The immerse-boundary thickness area in BDIM is $\epsilon/\Delta Y=2$, where $\Delta Y$ is cell-size. The area within $\epsilon$ is affected by the prescribed body velocities defining kinematics but still resolves the Navier-Stokes equations.

The foil is placed within a uniform Cartesian grid region with $2<\Delta X/\Delta Y<10$ and $2<\Delta Z/\Delta Y<8$, where $X,Y,Z$ are the global axes in inflow streamwise, lateral and spanwise of the domain. Outside the uniform grid, the grid domain stretches from $X=-3C$ to $X=11C$ and $Y=\pm 3C$. The size of the uniform grid is set such that the first grid points outside the foil are always inside the linear boundary layer region below or $y^+ \leq10$, which $y^+=y u_{\tau}/\nu$ is the viscous distance and $u_\tau$  friction velocity. The first grid points away from the foil reach the resolutions defined in detail in Tab. \ref{tab:resolution} after flapping cycles where the simulations converge. The $x,y,z$ are local tangential, normal, and spanwise distances respectively to the solid boundary as depicted in Fig. \ref{fig:phase}(a). The spanwise domain is uniform with periodic boundary conditions, and the domain span $S$ is set to be minimal, but the resolution is kept such that $S \ge 100 \nu/u_{\tau}$ to sufficiently capture the smallest structure in the spanwise direction based on \cite{Jimenez1991}. Each 3-D simulation was continued and expanded from the 2-D pre-simulation, which had converged over 15 cycles.

\begin{table}
  \begin{center}
  \begin{tabular}{cccccc}
    \toprule
      $Re$  		 &  resolution $C/\Delta Y$ & $y^+$ range 		&  Span length/C & 3-D simulation cycles & Phase-averaging cycles\\
      $10^4$   & 700 								 & $\approx [1 .. 5]$	& 0.5 				  & 70 			  & 70 \\
      $10^5$   & 5000 							 & $\approx [1 .. 5]$ & 0.1 				  & 67			  & 67 \\
      $10^6$  	 & 8555 							 & $\approx [3 .. 10]$ & 0.009 			  & 16			  & 6 \\
    \bottomrule
  \end{tabular}
  \caption{Resolution for three $Re$.The total cycles of 3-D simulations exclude the 2-D pre-simulations of 15 cycles. The phase-averaging is conducted at the last cycles of 3-D simulations. } 
  \label{tab:resolution}
  \end{center}
\end{table}

We discuss the comparison for $Re=10^4$, $10^5$ and $10^6$ for this study in Section \ref{sec:results}. We limit the $Re$ to one million due to the high computational cost of wall-resolved LES simulations and Tab. \ref{tab:resolution} shows that our near-wall resolution and number of cycles had to be reduced for that case compared to the lower $Re$. Later in Section \ref{sec:results}, we notice that the phase averaging at $Re=10^6$ for 6 cycles was insufficient to completely remove the fluctuations, and might need more resolution near the trailing edge region. However, this limitation does not change the conclusion of the study, especially the trends with increasing $Re$. Similarly, the lack of resolution affects only the flow far downstream.

\subsection{Cycle phases}

In the result discussion, we will present data for 32 phases per cycle as per Fig. \ref{fig:phase}(b). The results will only represent the upper side of the foil experiencing phases per cycle $t/T$s. Later, we show the span and phase averaging results as presented at Tab.\ref{tab:resolution} for the number of cycles for averaging. Since the foil is symmetrical, the lower side experiences the same states as the upper side after a half cycle or $t/T+0.5$. In the results, we present 32 phases per cycle and the first phase starts when the foil moves upwards $t/T=0.03125$ and ends at phase $t/T=1$  when the foil pivot at location $Y=0$. The cycle restarts after $t/T=1$ or is equivalent to $t/T=0$. The projected positions are presented in 100 $x/C$ equispaced locations on the foil as depicted in Figure \ref{fig:phase}(a). 

Later in the result section, we present parameters needed to compute the boundary layer edge velocity $u_e$. We use the method of velocity reconstruction from pressure data by \cite{Griffin2021}, and has been compared with other methods such as by \citet{Uzun2021}. This velocity reconstruction is more suitable for our cases because the velocity profiles expand and retract dynamically along the phases and locations as explained in \cite{Zurman2024}.

\section{Results} \label{sec:results}

\subsection{Comparison with experiments}\label{sub:compare_exp}

\begin{figure}[tbhp]
    \centerline{\includegraphics[width=0.7\textwidth]{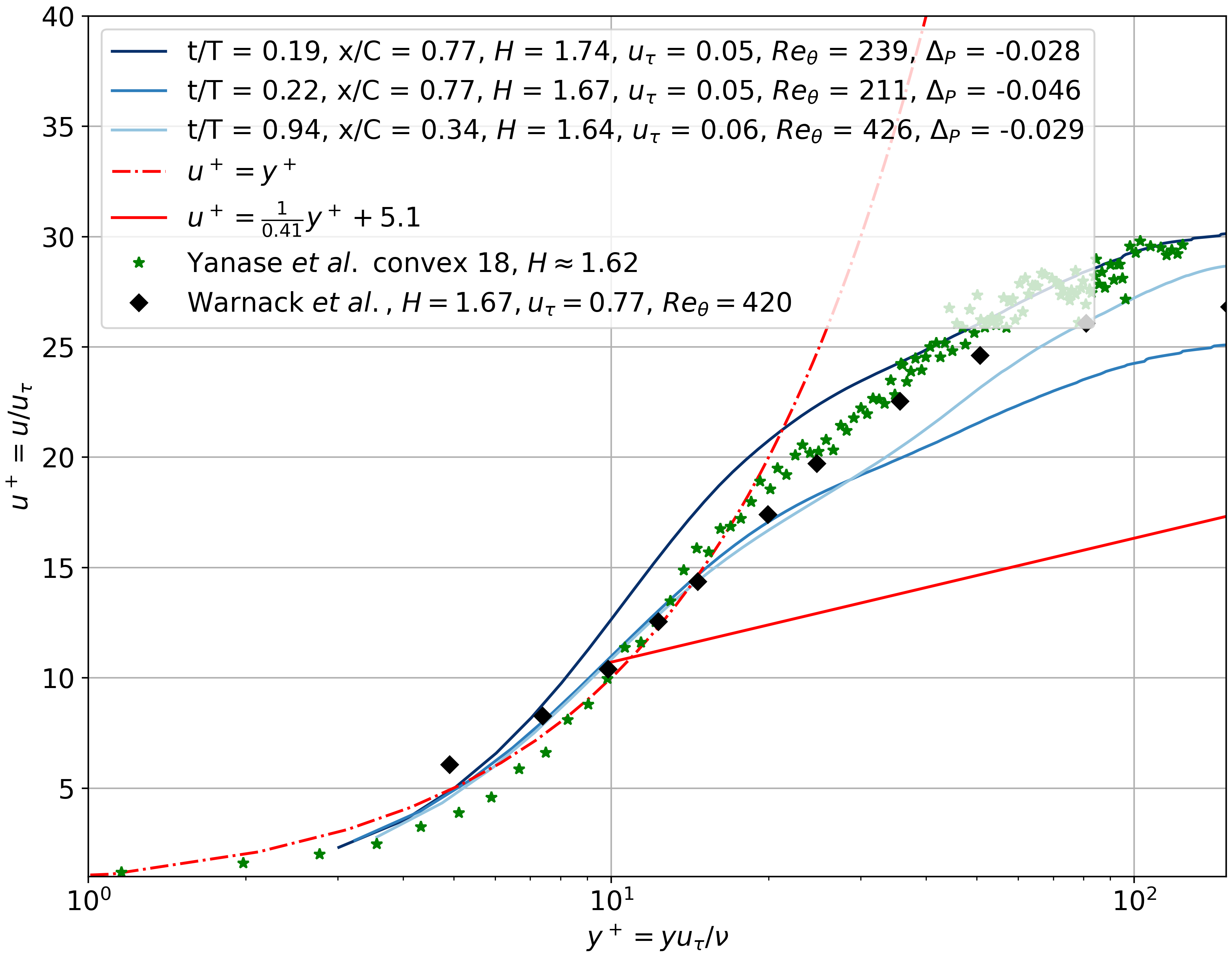}}
    \caption{Velocity profile comparison of the current study (blue straight lines) with experiments. The current study is at $Re=10^5$ at several $t/T$ phases and $x/C$ locations. Presented experiments are from the Rainbow trout experiment by \citep{Yanase2015} on convex-motion (unknown $u_{\tau}$ and $Re_{\theta}$), and the turbulent boundary layer at FPG by \citep{Warnack1998}. Despite having different $u_{\tau}$, the profiles show similar extended laminar region.}
        \label{fig:Experiment_comparison}
 \end{figure}

Despite the scarcity, we found comparison studies for our flapping foil studies at $Re \geq 10^5$, such as the experiment of swimming Rainbow Trout by \citet{Yanase2015} at body-length $Re_{BL} \approx 10^5$ and the study of FPG effects on axisymmetric turbulent boundary layers by \citet{Warnack1998}. By matching the boundary layer characteristics such as the shape factors $H=\delta^*/\theta$ (ratio of displacement $\delta^*$ to momentum thicknesses $\theta$), we recover similar velocity profiles as shown in Fig. \ref{fig:Experiment_comparison}. One sample in the current study is relatively close to the momentum-thickness Reynolds numbers $Re_{\theta}$ \citep{Warnack1998}, but the friction velocities $u_{\tau}$ are completely different. There is no information on $Re_{\theta}$ and $u_{\tau}$ from the experiments of swimming Rainbow Trout for comparison.

The obvious characteristic of the velocity profiles mentioned in the experiments and our current study is the extended laminar region beyond $y^+=11$. Most of the velocity profiles of our study show extended laminar regions as in Fig. \ref{fig:Experiment_comparison}. We hypothesize that velocity profiles of flapping foils and swimming animals at optimum propulsion might show similar profiles deviating from the log region found in turbulent flow over the flat plate at ZPG. Despite the shape factor being lower than $H<2.5$, the velocities show laminar profiles. The value of $H\approx 1.7$ is considered in the transition to turbulent and far lower than the laminar shape factor values mentioned by \citet{Schlichting1979}. This phenomenon is considered a \textit{relaminarization}, similarly mentioned by \citet{Warnack1998, Balin2021} at turbulent boundary layer under the influence of strong pressure gradients. In the case of a stationary foil at $Re=10^5$ and angle of attack lower than $\alpha_{eff}=23.8^{\circ}$, the flow would have experienced strong separation or stall-triggered turbulent flow as seen in the study of \citep{Vinuesa2018}. Flapping foils are known to experience dynamic stall which delays the separation. 

\citet{Patel1968} called the \textit{relaminarization} phenomenon a \textit{reverse transition} and suggested that its onset happens in the boundary layer mostly with a strong favourable pressure gradient, and it is independent of the boundary layer characteristics such as $H$ or $Re_{\theta}$. Apart from similar $H$, both $u_\tau$ and $Re_{\theta}$ in this study can be different, as shown in Fig. \ref{fig:Experiment_comparison}. However, the non-dimensionalised pressure gradient $\Delta_P$ is below the threshold of -0.018, indicating the influence of strong adverse gradients. \citet{Patel1968}[pp.376] shows a similar example where strong pressure gradients cause a major departure from the semi-logarithmic law, not only the adverse but also the favourable pressure gradients.  Since flapping foils are highly affected by the pressure gradient, as we will discuss further in the next section, this should be the mechanism explaining the \textit{relaminarization} boundary layer profile. In the following sections, we will discuss the effect of the pressure gradient originating from the outer part of the boundary layer.
 
\subsection{Freestream region influence}  

\begin{figure}[tbhp]
	\centering
	\begin{subfigure}[b]{0.49\textwidth}
    	\includegraphics[width=1\textwidth]{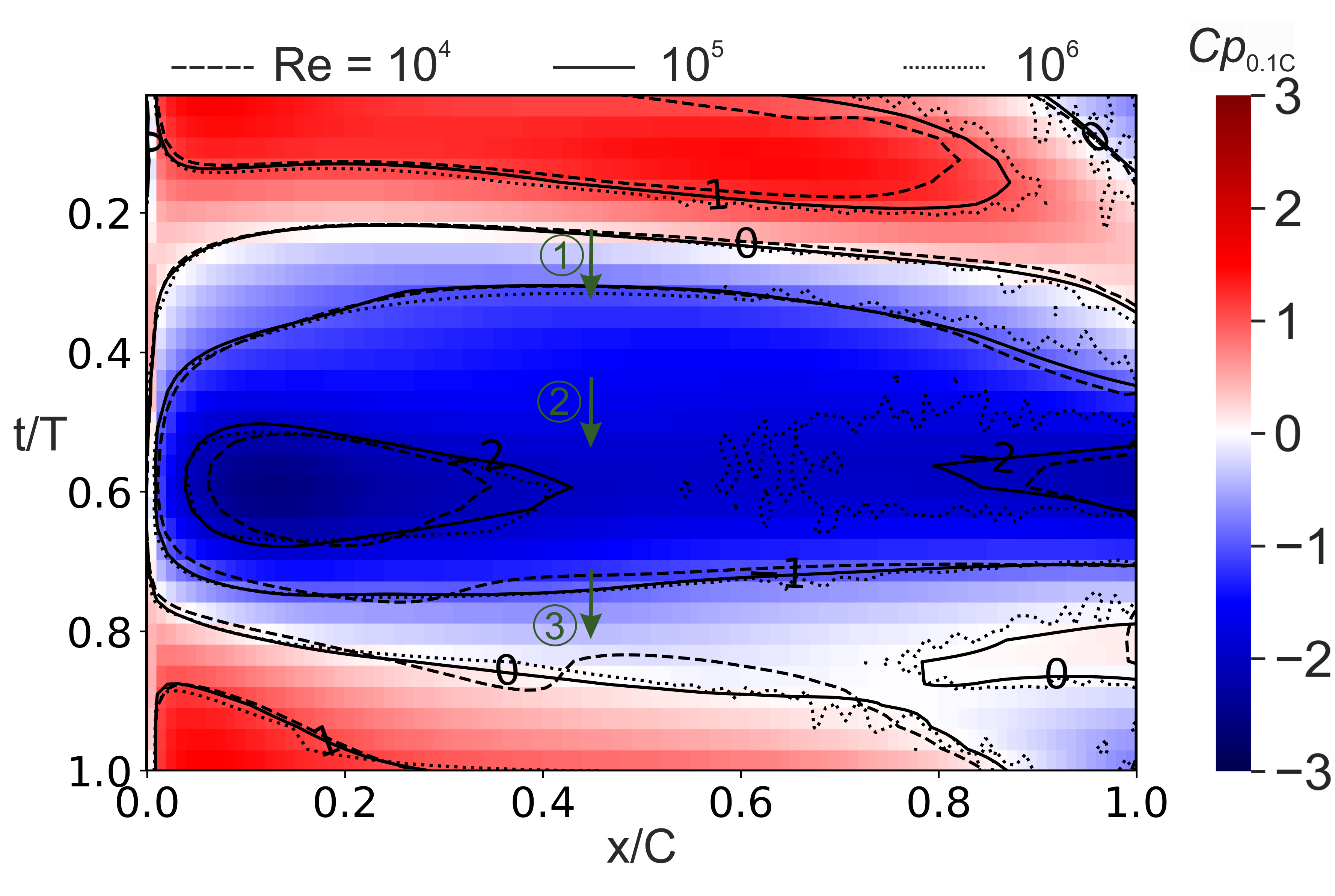}
        \subcaption{}
	\end{subfigure}
	\begin{subfigure}[b]{0.49\textwidth}
    	\includegraphics[width=1\textwidth]{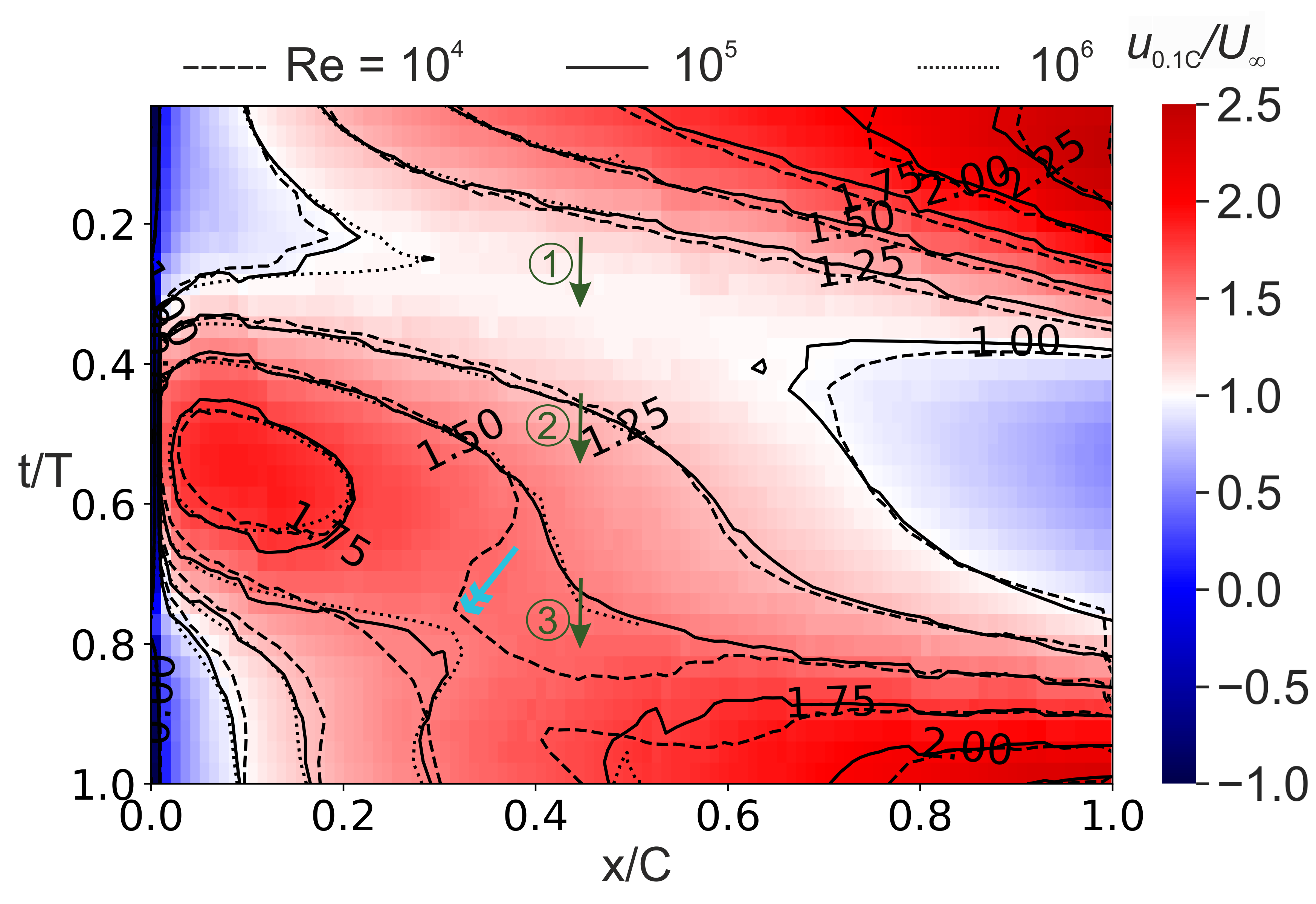}
        \subcaption{}
	\end{subfigure}
	\begin{subfigure}[b]{0.51\textwidth}
    	\includegraphics[width=1\textwidth]{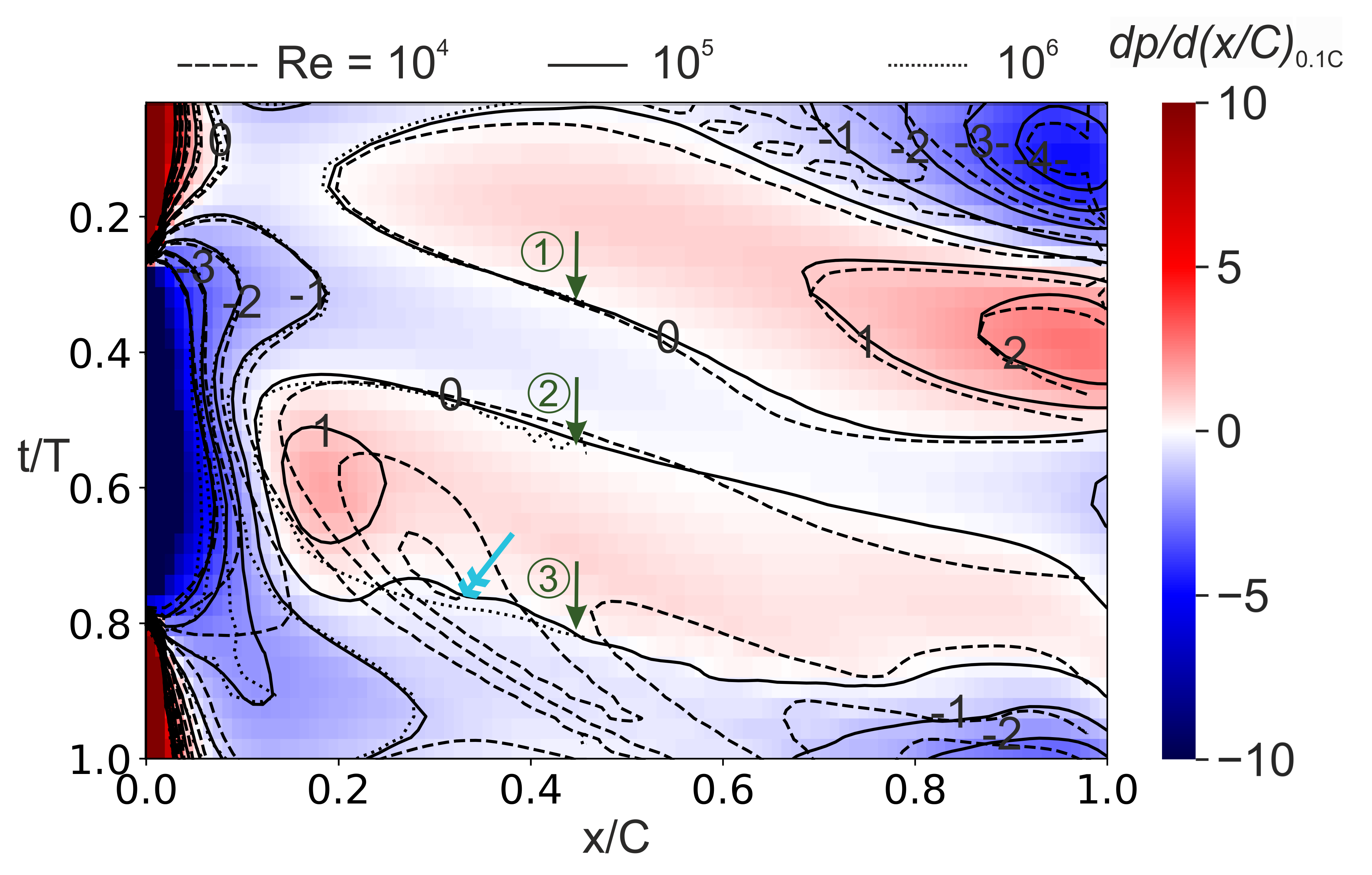}
        \subcaption{}
	\end{subfigure}
	\caption{The heatmaps of (a) pressure coefficients ($Cp$), (b) streamwise local velocities $u/U_{\infty}$, and (c) pressure gradient $\dv*{p}{(x/C)}$ at the height of $0.1C$ from the wall at $Re=10^5$. The horizontal axis is the projected position on the upper foil $x/C$ representing space, and phase $t/T$ on the vertical axis represents the time evolution of the flow. All are span- and phase-averaged data with selected contour lines to compare different $Re$. At $0.1C$ height, the heatmap represents the freestream region. Double-head arrows (cyan) show the deviated contour at $Re=10^4$ due to an LEV. Green arrows with numbers are given for later discussion in Fig.\ref{fig:velo_profiles}. }
	\label{fig:freestream_parameters}
 \end{figure}  

Fig. \ref{fig:freestream_parameters} shows the pressure and velocity variation at a height of $y=0.1C$ above the foil as a function of length along the foil and phase in the motion cycle across the three Reynolds numbers. The pressure coefficient heatmap in Fig. \ref{fig:freestream_parameters}(a) shows that the suction region (negative $Cp$ values) occurs for phases $ 0.3 \leq t/T \leq 0.7$ and is fairly uniform from the leading to the trailing edges. Each side of the foil experiences alternately pressure side or suction side once at a time per cycle. 

In Fig. \ref{fig:freestream_parameters}(b), the local velocities $u/U_{\infty}$ at height $0.1C$ show that each foil side also experiences alternating high (a peak) or low (a valley), but they are not uniform along the chord length. Each foil side can experience different velocities in one phase. For example at phase $t/T\approx 0.5$, the leading edge experiences a much higher velocity than the inflow with $u > 1.75U_{\infty}$, while the trailing edge experiences $u < 0.5U_{\infty}$. The velocity countours show an inclination of about a half cycle ($0.5T$) from the leading edge at $t/T \approx 0.5$ to the trailing edge $t/T \approx 1$. This half-cycle inclination is due to the sinusoidal flow advection from the prescribed flapping kinematics. 

We also present the pressure gradient evolution $\dv*{p}{(x/C)}$ in Fig. \ref{fig:freestream_parameters}(c) at height $0.1C$. Each position of $x/C$ alternates \textit{twice} between adverse and favorable pressure gradients per cycle with two distinct APG regions, except the leading edge region, and the trailing edge which is slightly clipped by the separated wake. The trend of APG and FPG is inclined similarly to the velocity, following the flow advection speed.

At height $0.1C$, increasing Reynolds numbers do not show a significant difference, as shown by the majority of the contour lines of different $Re$. There is data noise for $Re=10^6$, especially near the trailing edge, due to the least number of cycles for phase averaging. We can see that the variation is very minimal for all $Re$ with an exceptional deviation caused by a single leading-edge vortex affecting the $y=0.1C$ region of $Re=10^4$. At around $t/T=0.75$ and $x/C=0.4$, a double-headed cyan arrow in Fig. \ref{fig:freestream_parameters}(b or c) shows an elongated region that takes a longer phase $t/T$ to disperse the flow downstream. It is caused by an LEV that is advected more slowly than the half-cycle advection due to its large size. We discuss further in the next section how larger vortices tend to be advected more slowly than small ones.

\subsection{Vortex structures}

\begin{figure}[tbhp]
  \begin{tabular}{cc}
	\parbox{6.2cm}{\includegraphics[width=0.39\textwidth]{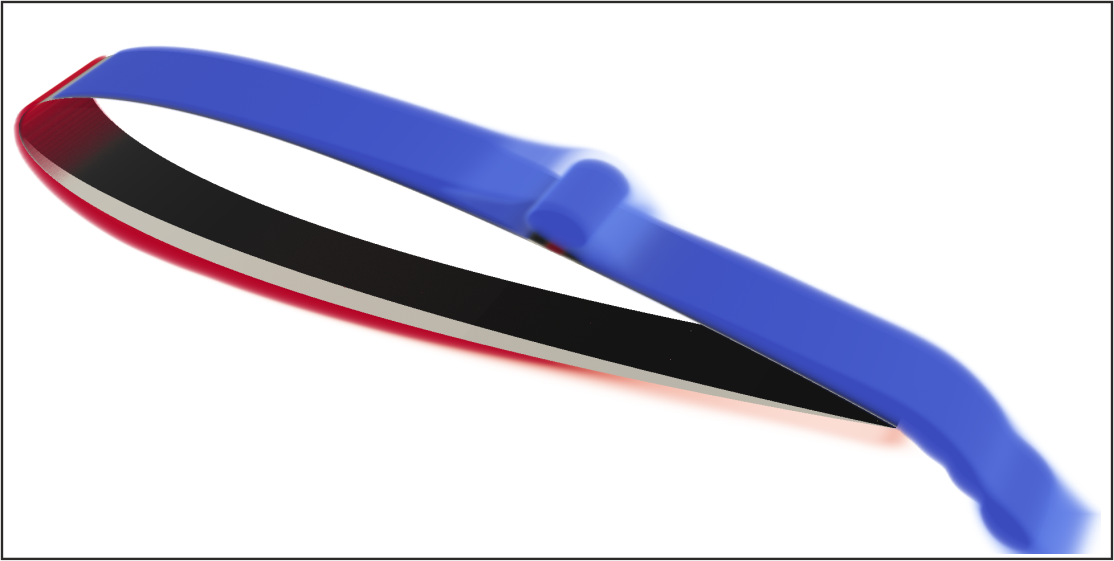}\\ 
			\centering (a)\\
			\includegraphics[width=0.39\textwidth]{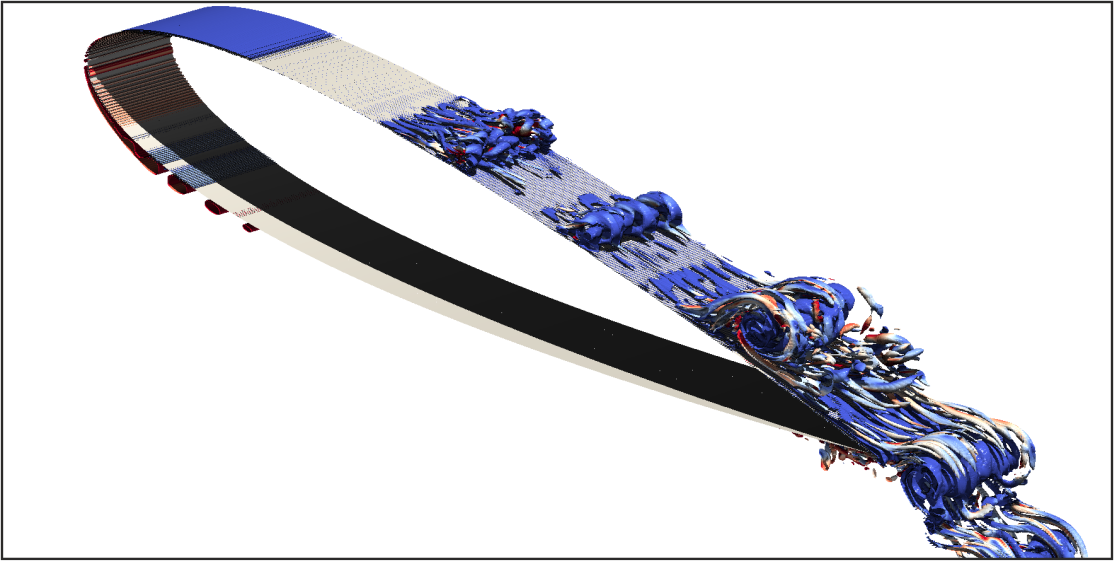} \\
			\centering (b)\\ } &
	\parbox{7cm}{\includegraphics[width=0.57\textwidth]{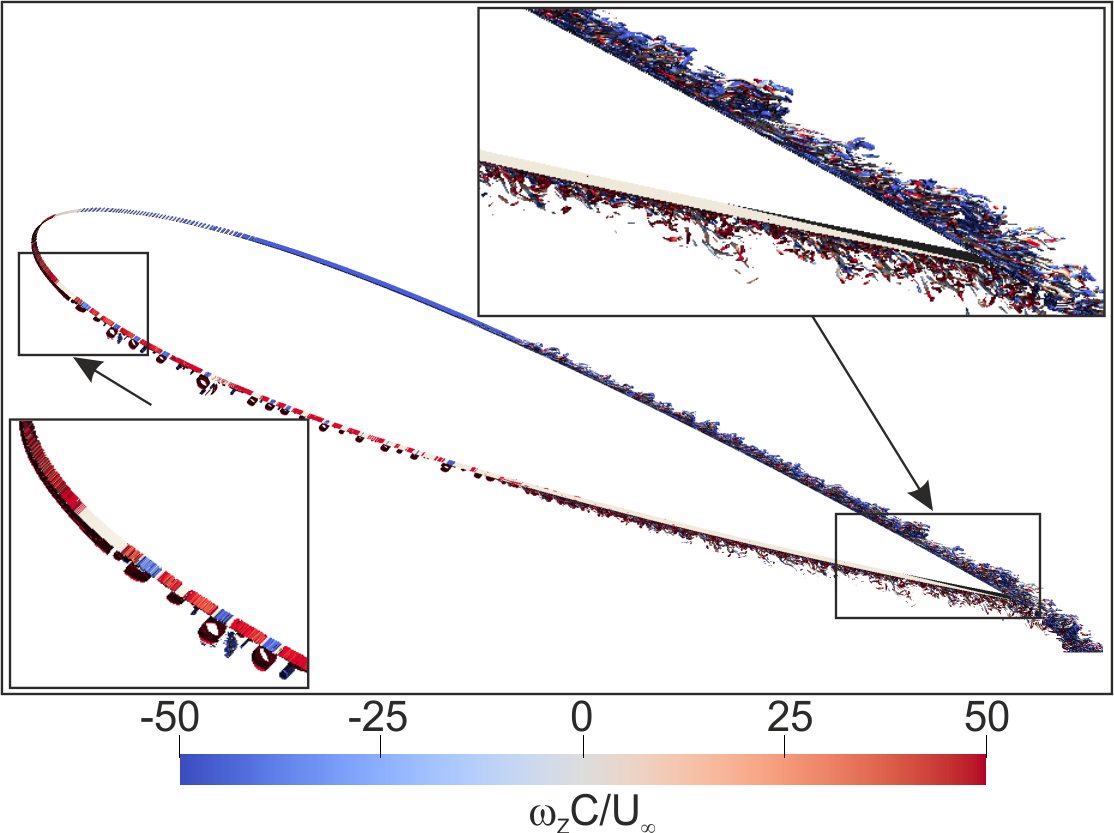}\\ 
			\centering (c)\\ }\\
  \end{tabular}
 \caption{Instantaneous vortex structures are represented by (a) volume render of spanwise vorticity of $Re=10^4$, (b) contour of 0.1\% maximum $\lambda_2$ of $Re=10^5$, and (c) $Re=10^6$. The upper foil experience phase $t/T=1$ or the last cycle, while the lower foil experience $t/T=0.5$. Colours represent spanwise vorticity normalised by the chord and the freestream velocity.} 
\label{fig:vorts}
\end{figure}

Increasing Reynolds numbers produce very distinct vortex structures generated by flapping foils, especially the LEVs. Fig.\ref{fig:vorts} presents the instantaneous vortex structures at $t=T$. The vorticity at $Re=10^4$ is essentially two-dimensional, Fig.\ref{fig:vorts}(a), matching the results from previous low $Re$ studies in the optimal  $0.15 \leq St \leq 0.45$ range \citep{Zurman2020}. A single large LEV affects the flow even at the height of $0.1C$ as previously pointed out by double-head cyan arrows in Fig. \ref{fig:freestream_parameters}(b and d). 

This two-dimensional flow cannot be maintained at higher $Re$, as seen in Fig.\ref{fig:vorts}(b,c). The flapping foil generates LEVs near the leading edge during the suction phase of each cycle, alternately on the upper or lower side. There is a single LEV at $Re=10^4$, but the flow produces smaller vortices and in much greater quantities as the $Re$ increases. These vortices are advected downstream and become unstable as the boundary layer grows, evolving into fully three-dimensional vortex structures around $0.4$ to $0.5C$ along the chord. As expected, the flow at $Re=10^6$ is noisier and more chaotic than at lower $Re$, especially near the trailing edge.

\subsection{The phase evolution of velocity profiles and near-wall pressure gradients}  

\begin{figure}[tbhp]
	\centering
	\begin{subfigure}[b]{1\textwidth}
    	\includegraphics[width=1\textwidth]{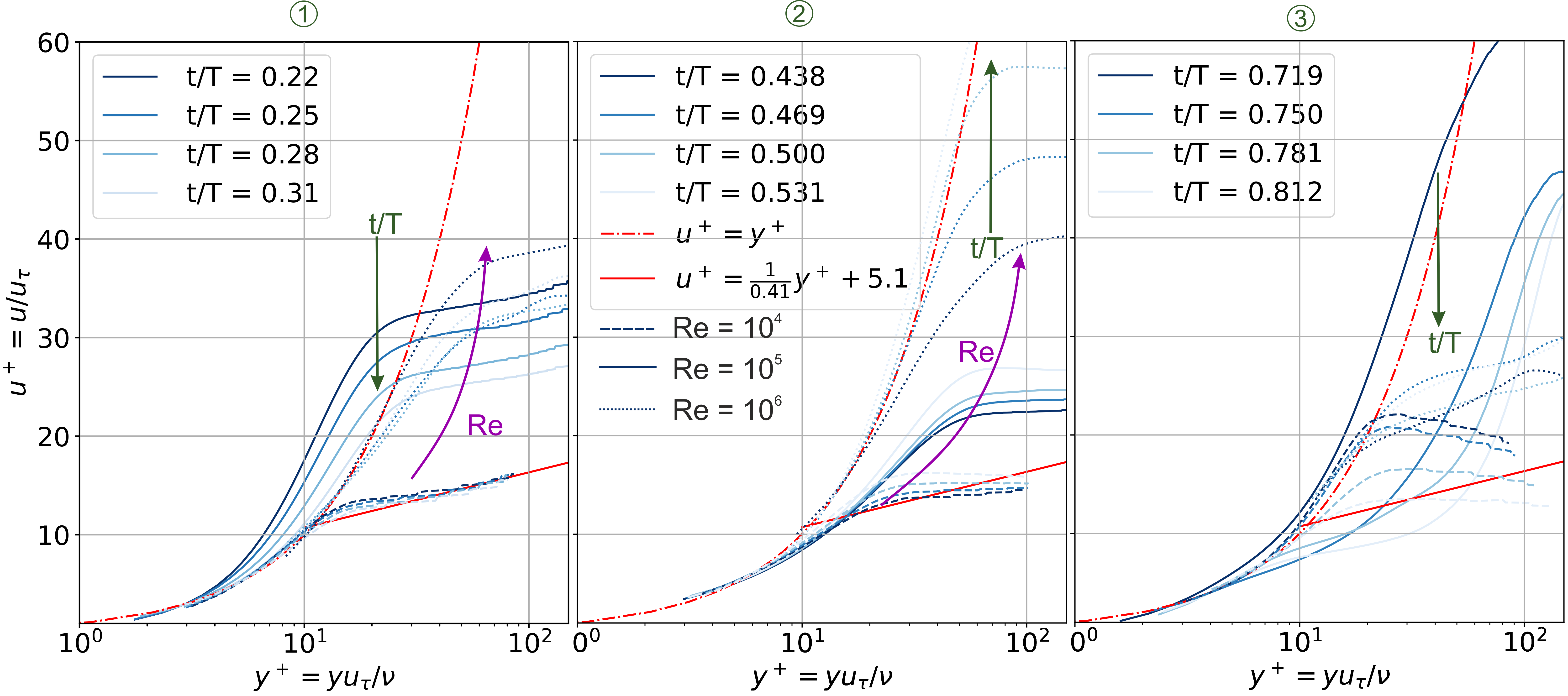}
	    \subcaption{}
	\end{subfigure}
	\begin{subfigure}[b]{1\textwidth}
    	\includegraphics[width=1\textwidth]{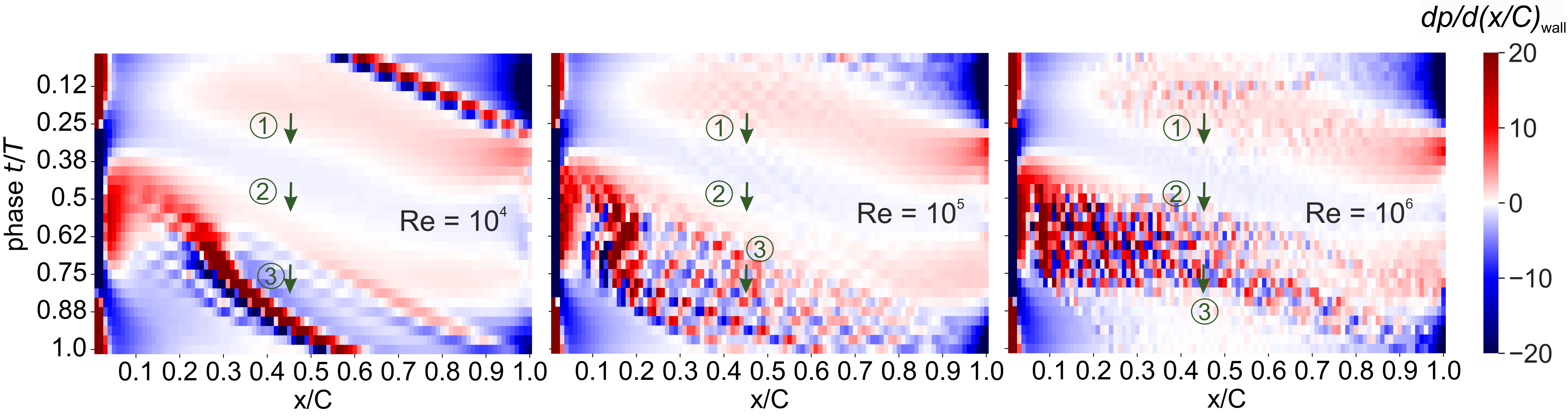}
	    \subcaption{}
	\end{subfigure}
	\caption{(a) Evolution of velocity profiles when exiting APG \textcolor{OliveGreen}{\protect\textcircled{\raisebox{1.pt} {\tiny 1}}}, exiting FPG \textcolor{OliveGreen}{\protect\textcircled{\raisebox{1.pt} {\tiny 2}}}, and inside vortex breakdown region \textcolor{OliveGreen}{\protect\textcircled{\raisebox{1.pt} {\tiny 3}}} at $x/C=0.45$. (b) Heatmaps of the near-wall pressure gradient for increasing $Re$.  Green arrows follow the direction of increasing phase $t/T$, and the magenta arrow shows the trend of increasing $Re$. Velocity profiles and wall pressure gradients $\dv*{p}{(x/C)}_{wall}$ are span and phase averaged.} 
\label{fig:velo_profiles}
\end{figure}

We found that the extended laminar profile at each location on the foil or the \textit{relaminarization} described in Section \ref{sub:compare_exp} earlier is promoted by the FPG, whereas APG reverses it. The green arrows on the freestream pressure gradient in Fig. \ref{fig:freestream_parameters}(c) can be compared to the pressure gradient near the wall in Fig. \ref{fig:velo_profiles}(b). The freestream region outside of the boundary layer acts as forcing, as marked by \textcolor{OliveGreen}{\textcircled{\raisebox{1pt} {\tiny 2}}}, accelerating the flow inside the boundary layer. 
This FPG acceleration causes the velocity profile to rapidly inflate the laminar-like region of the boundary layer, as shown in Fig. \ref{fig:velo_profiles}(a) \textcolor{OliveGreen}{\textcircled{\raisebox{1pt} {\tiny 2}}}. The whole boundary layer might not inflate, as the FPG tends to reduce the boundary layer momentum or displacement thicknesses. This expanding laminar part of the boundary layer is also more prominent with the increase of $Re$, with the $Re=10^6$ boundary layer tripling in size from $t/T=0.43-0.53$. Hence, the \textit{relaminarization} is stronger at higher $Re$.

On the contrary, the APG reverses the expansion of the laminar region. When the flow is at the APG \textit{relaminarization} process is reversed as in Fig. \ref{fig:velo_profiles} marked by \textcolor{OliveGreen}{\textcircled{\raisebox{1pt} {\tiny 1}}} and also \textcolor{OliveGreen}{\textcircled{\raisebox{1pt} {\tiny 3}}}. The extended laminar regions shrink during this phase as they move away from the linear curve $u^+=y^+$ and closer to the log profile. However, the shrinkage order based on increasing $Re$ at \textcolor{OliveGreen}{\textcircled{\raisebox{1pt} {\tiny 3}}} no longer follows the same trend as in \textcolor{OliveGreen}{\textcircled{\raisebox{1pt} {\tiny 1}}}. This is because the advection at \textcolor{OliveGreen}{\textcircled{\raisebox{1pt} {\tiny 3}}} of vortex breakdown now influences the flow in addition to the APG condition. 

Heatmaps in Fig. \ref{fig:velo_profiles}(b) show that pressure gradients at the wall differ for each $Re$. The breakdown of vortices dictates the pressure gradient distribution, as indicated by the flickering blue-red areas just after phase $t/T=0.5$. As LEVs at higher $Re$ are smaller, their vortices experience faster advection. The arrow marked by \textcolor{OliveGreen}{\textcircled{\raisebox{1pt} {\tiny 3}}} is at the onset of LEV advection for $Re=10^4$, while it is within the advection region of LEV breakdown at $Re=10^5$ and has just passed through at $Re=10^6$. The reverse effect of \textit{relaminarization} ends faster at higher $Re$ due to faster advection. 

\subsection{Indication of \textit{relaminarization} and cyclic behaviour}  

\begin{figure}[tbhp]
	\centering
        \includegraphics[width=1\textwidth]{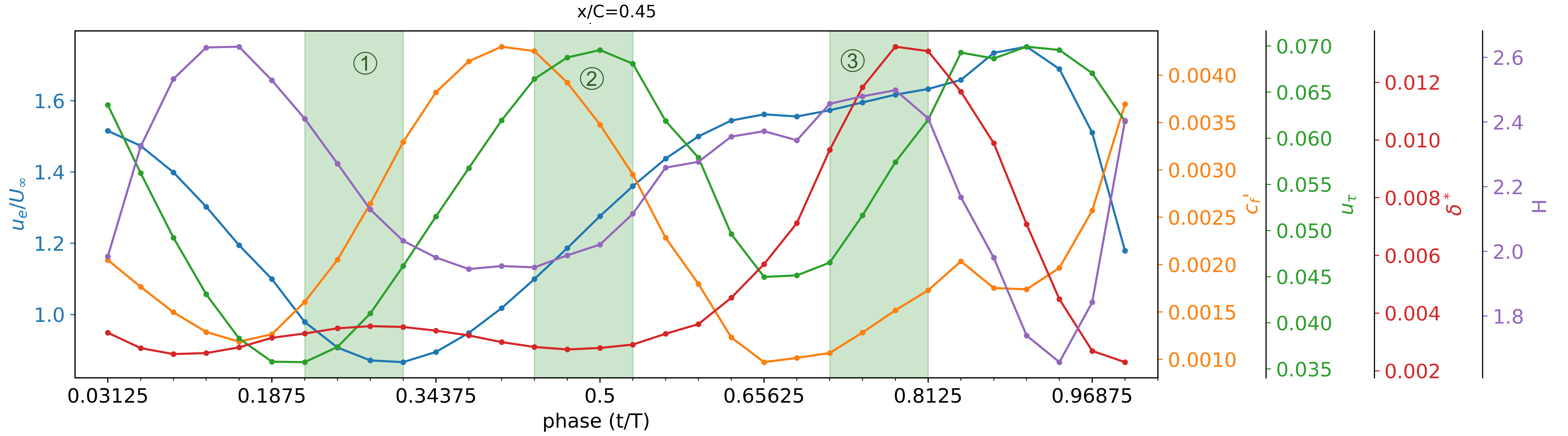}
	\caption{Evolution of variables per cycle at $Re=10^5$ (values are span and phase-averaged). Shaded areas correspond to green arrows in Fig. \ref{fig:velo_profiles}.}
	\label{fig:innerBL_parameters}
 \end{figure} 

Fig.\ref{fig:innerBL_parameters} illustrates the cyclic evolution of the normalised boundary layer edge velocity $u_e/U_{\infty}$, the zero-mean fluctuation of wall-friction coefficient $c_f'$, friction velocity $u_{\tau}$, displacement thickness $\delta^*$, and shape factor $H$ at about mid position $x=0.45C$. The $c_f'$ is the zero-mean fluctuation of $c_f=2(u_{\tau}/u_e)^2$, where the average across phase $t/T$ is removed at each position $x/C$. During the expansion of the laminar region marked by \textcolor{OliveGreen}{\textcircled{\raisebox{1pt} {\tiny 2}}}, same as in Fig. \ref{fig:freestream_parameters}(b) and \ref{fig:velo_profiles}, the freestream velocity is positive and the flow is accelerating or at FPG state. Region \textcolor{OliveGreen}{\textcircled{\raisebox{1pt} {\tiny 2}}} begins with the peak of $c_f'$, followed by a rapid decrease by 31\% for an eighth of a cycle. \citet{Uzun2021, Mukund2006} found a similar trend of decrease. This condition is accompanied by an 8.5\% increase in $H$, explaining the expansion of the laminar-like region. On the contrary, the APG areas of \textcolor{OliveGreen}{\textcircled{\raisebox{1pt} {\tiny 1}}} and \textcolor{OliveGreen}{\textcircled{\raisebox{1pt} {\tiny 3}}} indicate an increase of $c_f'$ by double and 62\%, along with the decrease of $H$ by 15\% and 2\%, explaining the shrinkage of the laminar-like region of the boundary layer. Friction velocity $u_{\tau}$ follows the same trend as $c_f'$ with a slight delay. The $\delta^*$ is at its lowest value during FPG with a sudden increase by 35\% occurring at \textcolor{OliveGreen}{\textcircled{\raisebox{1pt} {\tiny 3}}} mostly due to APG region and downstream vortices. Both boundary layer displacement $\delta^*$ and momentum thicknesses $\theta$ (not shown) show a similar trend. 

\begin{figure}[tbhp]
	\centering
        \includegraphics[width=1\textwidth]{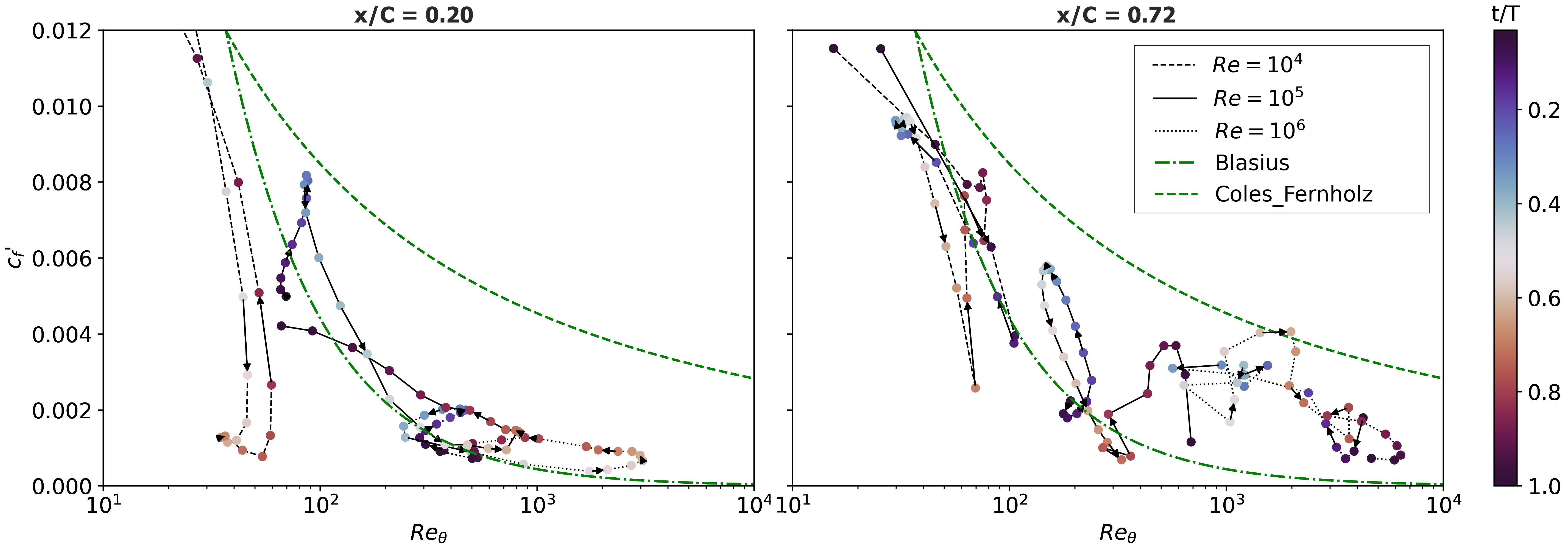}
	\caption{Cyclic behavior of $c_f'$ at $x/C=0.2$ and $0.72$ for different $Re$. Correlation lines are shown by the Blasius (laminar) and Coles \& Fernholz (turbulent). Arrows and colors (legend) show forward in phase $t/T$. Low $Re$ cycles live closer to laminar, but the higher $Re$ cycles intermittently switch between laminar and turbulent.}
	\label{fig:cf_prime_cycle}
 \end{figure} 

Graphs in Fig. \ref{fig:cf_prime_cycle} demonstrate that each location $x/C$ exhibits a cyclic behaviour of the friction coefficient fluctuation $c_f'$ that intermittently approaches either laminar or turbulent limits. Here, the laminar correlation is presented by Blasius line where $c_f=(0.664^2)/Re_{\theta}$, and turbulent at ZPG by Coles \& Fernholz correlation where $c_f=2 \left( \frac{1}{0.41} \log (Re_{\theta})+4.127 \right) ^{-2}$. Increasing global $Re$ also increases $Re_{\theta}$, shifting the cycles to the left side of the graph. However, each chordwise location exhibits different cyclic behaviors. Near the upstream locations ($x/C=0.2$), the $c_f'$ cycles are closely aligned with the Blasius or laminar correlation limit even for $Re=10^6$. Moving closer to the trailing edge at $x/C=0.72$, they begin to approach the turbulent correlation line, especially at higher $Re$. Fig. \ref{fig:cf_prime_cycle} indicates that most cycles at $Re=10^4$ stay laminar, while $c_f'$ at higher $Re$ indicate a cyclic behaviour between laminar and turbulent. 

\subsection{Failure of the standard \textit{relaminarization} criteria}

\begin{figure}[tbhp]
	\centering
        \begin{subfigure}[b]{.30\textwidth}
    	\includegraphics[width=1\textwidth]{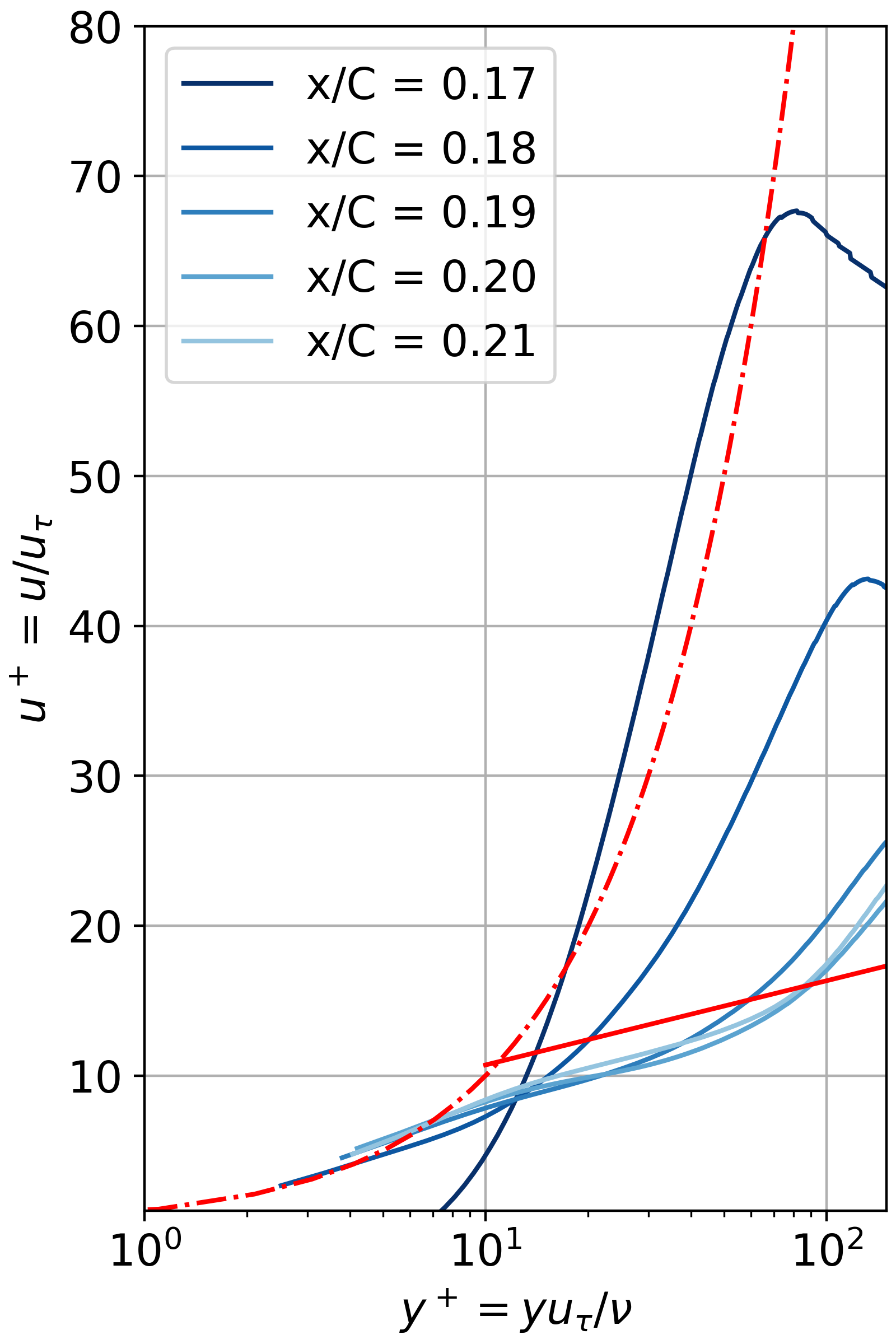}
	    \subcaption{}
	\end{subfigure}
        \begin{subfigure}[b]{.285\textwidth}
    	\includegraphics[width=1\textwidth]{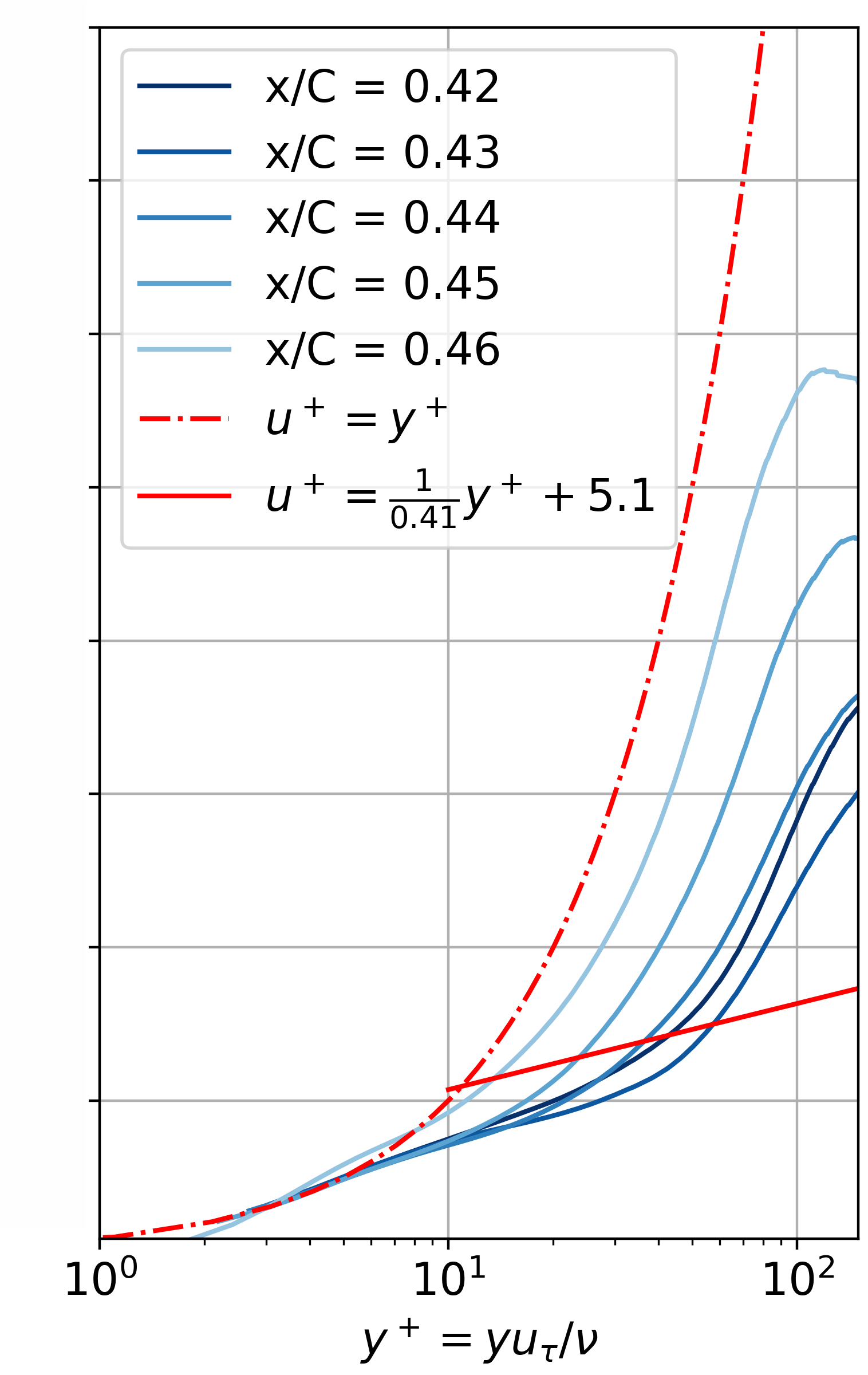}
	    \subcaption{}
	\end{subfigure}
        \qquad
        \begin{tabular}[b]{ccccccc}
          \toprule
          $x/C$ & $H$ & $Re_{\theta}$ & $\Delta_p$ & $\Delta_{\tau}$ & $K$ & $c_f'$\\\midrule
          0.17 & 3.46 & 486 & -7.68 &$1.0\times10^{-5}$&$4.5\times10^{-6}$& $4.9\times10^{-4}$\\
          0.18 & 2.35 & 668 & -1.29 &$-1.0\times10^{-3}$&$-1\times10^{-5}$& $1.2\times10^{-3}$\\
          0.19 & 1.90 & 713 & -0.19 &$-6.6\times10^{-4}$&$-1.3\times10^{-5}$& $2.9\times10^{-3}$\\
          0.20 & 1.78 & 707 & 0.047 &$-5.2\times10^{-4}$&$-5.8\times10^{-6}$& $3.6\times10^{-3}$\\
          0.21 & 1.78 & 716 & 0.050 &$-4.9\times10^{-4}$&$-1.3\times10^{-6}$& $3.4\times10^{-3}$\\
          \midrule
          0.42 & 2.28 & 733 & -0.67 &$-6.5\times10^{-4}$&$-1.5\times10^{-5}$& $1.8\times10^{-3}$\\
          0.43 & 2.20 & 725 & -0.10 &$-8.1\times10^{-4}$&$4.5\times10^{-6}$& $2.3\times10^{-3}$\\
          0.44 & 2.31 & 729 & 0.383 &$-7.5\times10^{-4}$&$1.2\times10^{-5}$& $1.7\times10^{-3}$\\
          0.45 & 2.51 & 742 & 0.704 &$-3.7\times10^{-4}$&$1.2\times10^{-5}$& $1.1\times10^{-3}$\\
          0.46 & 2.60 & 758 & -0.80 &$6.7\times10^{-4}$&$5.1\times10^{-6}$& $7.0\times10^{-4}$\\
          \bottomrule
        \end{tabular}
	\caption{(a) Decelerating and (b) accelerating flow (\textit{relaminarization}) at phase $t/T=0.75$ at $Re=10^5$ are observed for the deviation from standard log law. The variable nominals are indicated in the table. }
	\label{fig:laminar_transition}
 \end{figure} 

Several criteria for \textit{relaminarization} or flow reversion, as mentioned by \citet{Narasimha1978} from various studies, show that the deviations in the velocity profile from the standard log law are inconsistent for flapping foils. This inconsistency arises due to the influences from upstream conditions and the preceding phases of the flow. Criteria for \textit{relaminarization} occur when $K=\frac{\nu}{\rho u_e^2} \frac{du_e}{dx} \gtrsim 3.5 \times 10^{-6}$, $\Delta_p =-\frac{\nu}{\rho u_{\tau}^3} \left.\frac{dp}{dx}\right\vert_{wall} \gtrsim 0.025$ and  $\Delta_{\tau} =\left.\frac{\nu}{u_{\tau}^3}\frac{d \tau}{dy}\right\vert_{wall} \lesssim 0.009$. 

Fig. \ref{fig:laminar_transition} illustrates the velocity profile evolution on sequential chord-wise positions in one phase and the alignment with the standard log law. Fig. \ref{fig:laminar_transition}(a) shows a decelerating flow where we observe a deviation of velocity profiles from the log law at positions $0.17\leq x/C \leq 0.18$, while $0.19\leq x/C \leq 0.21$ revert to the log law. However, $K$ below its criterion starts from $x/C \geq 0.18$, earlier than the change observed in velocity profiles. $\Delta_{\tau}$ values fall below zero while $\Delta_p$ increases, indicating APG, but they do not align with both criteria. Fig. \ref{fig:laminar_transition}(b) shows an accelerating flow where we expect \textit{relaminarization} and deviation from the log law has started at $x/C \geq 0.42$. $K$ increasing beyond its criterion starts late at location $x/C \geq 0.43$, $\Delta_p$ above criterion only occurs at $0.44\leq x/C \leq 0.45$, and $\Delta_{\tau}$ are all below the criterion. 

The inconsistencies with flow reversion limits mentioned in the literature might be attributed to the requirement of a constant pressure gradient and the absence of flow history from the previous phase and upstream flow. Subtracting the influence from upstream flow or previous phases requires a more controlled condition, which is beyond the current studies. The obvious indications of the log-law departure for the flapping foil observed in this study occur when the friction coefficient $c_f'$ decreases while the shape factor increases above the transitional turbulence $H \gtrsim 1.8$, as mentioned in the previous section. A more controlled study is necessary in the future to confirm the precise limits.
    
\subsection{Wall friction fluctuation and Clauser parameter} 
The previous sections provide a clear indication that both the shrinkage and contraction of the laminar-like regions are directly related to the wall friction coefficient fluctuation ($c_f'$). Therefore, it is essential to examine how it behaves with increasing $Re$.

\begin{figure}[tbhp]
	\begin{subfigure}[b]{0.48\textwidth}
    	\includegraphics[width=1\textwidth]{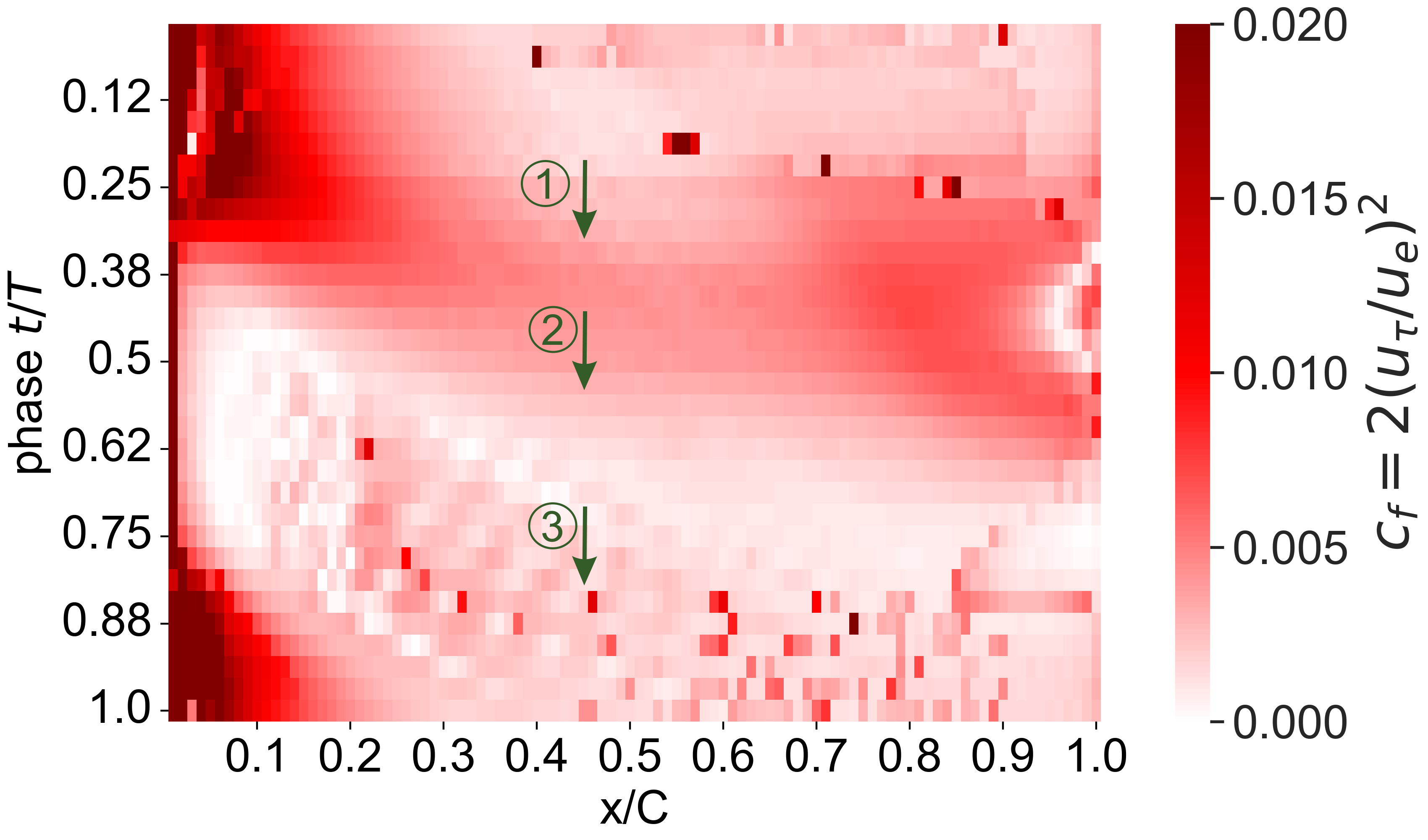}
        \subcaption{}
	\end{subfigure}
	\begin{subfigure}[b]{0.48\textwidth}
    	\includegraphics[width=1\textwidth]{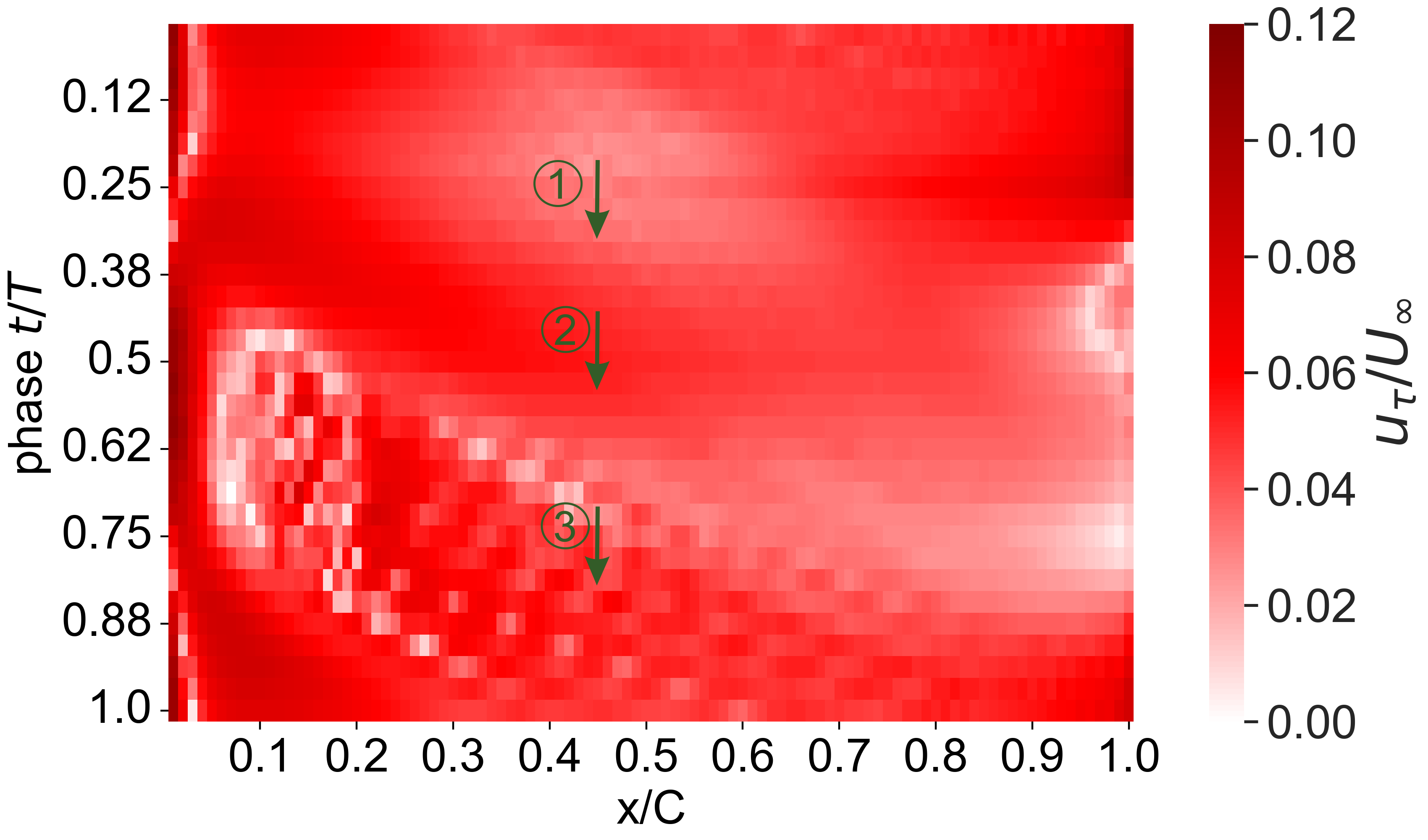}
        \subcaption{}
        \end{subfigure}
	\begin{subfigure}[b]{0.48\textwidth}
    	\includegraphics[width=1\textwidth]{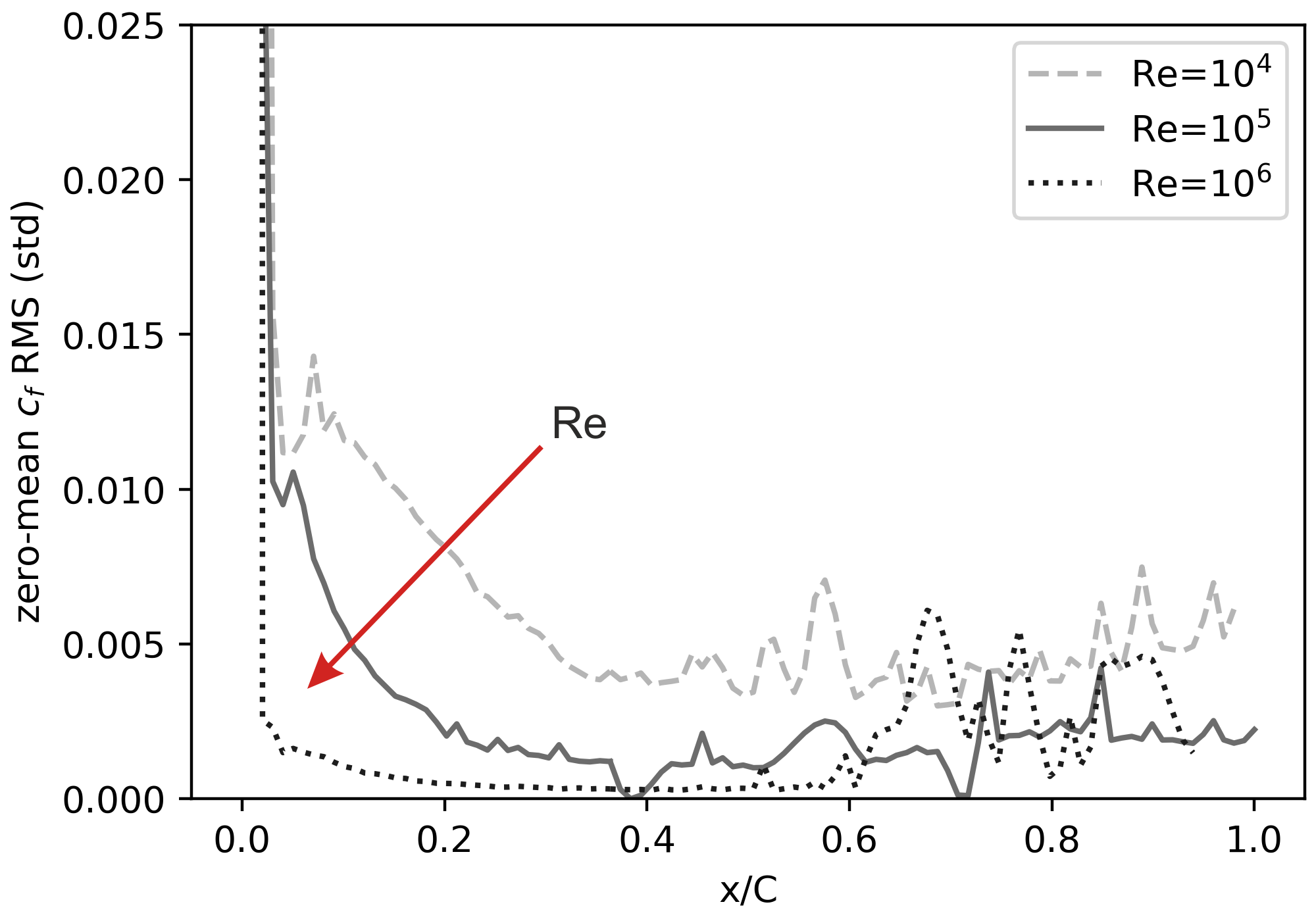}
        \subcaption{}
	\end{subfigure}
	\begin{subfigure}[b]{0.48\textwidth}
    	\includegraphics[width=1\textwidth]{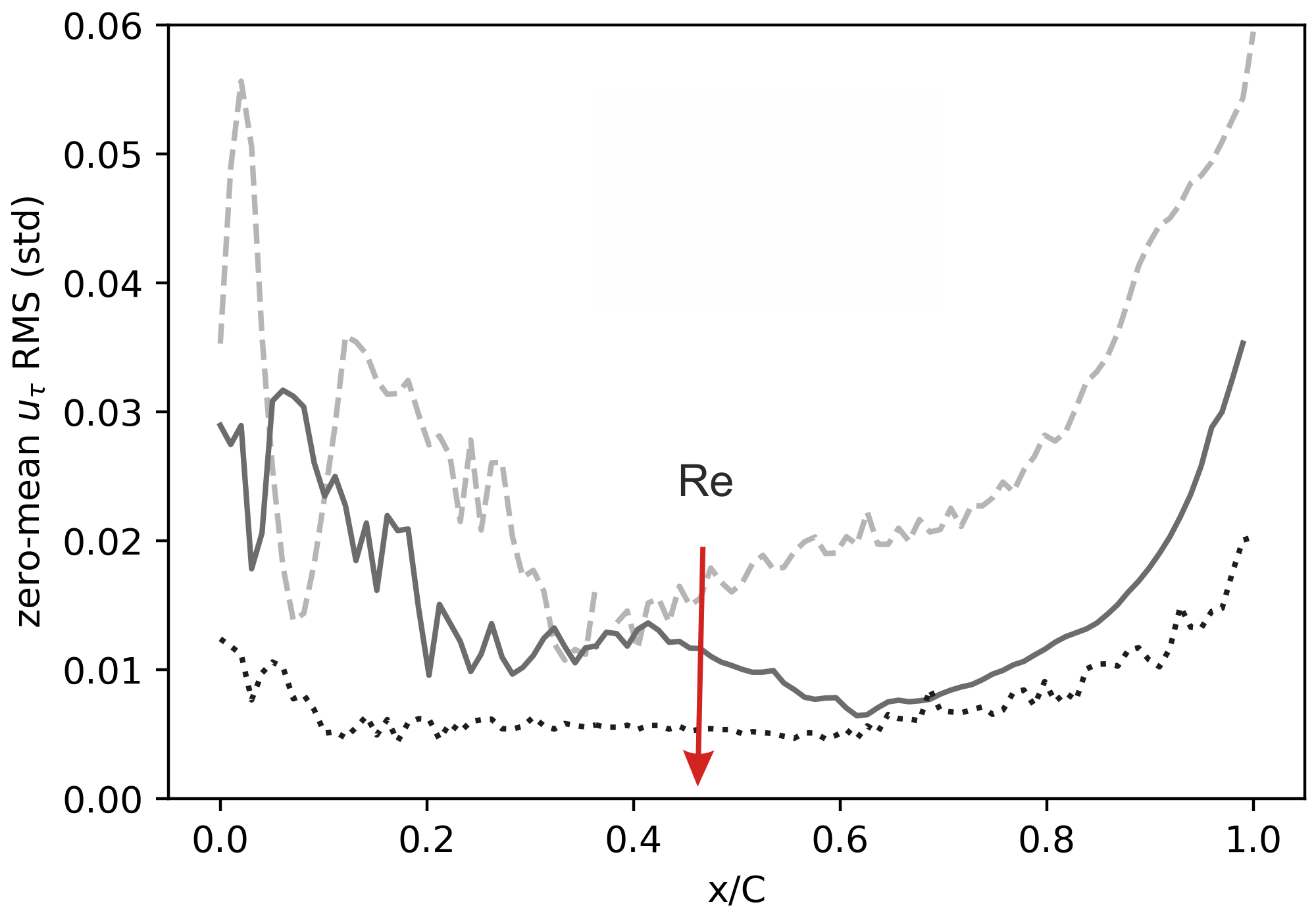}
        \subcaption{}
	\end{subfigure}
	\caption{Heatmap of (a) the wall friction coefficient $c_f$ and (b) friction velocities $u_{\tau}$ of $Re=10^5$ along the streamwise location and phase. The zero-mean Root-Mean Square (RMS or Standard Deviation) of (c) $c_f$ and (d) $u_{\tau}$ with increasing $Re$ represents the fluctuation along the phases. $c_f$ and $u_{\tau}$ are span and phase averaged.}
	\label{fig:fluctuations}
 \end{figure} 

Fig.\ref{fig:fluctuations} (a,b) are heatmaps of the friction coefficient $c_f$ and the friction velocity $u_{\tau}$ variables for $Re=10^5$, where the trace of vortex breakdown is seen at the phase region $0.5 \lesssim t/T \lesssim 1$. Similar patterns of vortex advection are observed at other Reynolds numbers (refer to Appendix \ref{appA}), although each $Re$ exhibits distinct fluctuation distributions. The lowest values, depicted in the lightest color, are primarily observed in the region of LEV breakdown and its advection towards the trailing edge. Lower $c_f$ values are associated with turbulence.

We obtain the cyclic fluctuation across a cycle $T$ for $c_f$ and $u_{\tau}$, by zeroing the mean values at each $x/C$ from heatmaps in Fig.\ref{fig:fluctuations} (a,b). To compare fluctuations across different $Re$, we compute the standard deviation of $c_f$ and $u_{\tau}$, or zero-mean Root-Mean-Square (RMS) in Fig.\ref{fig:fluctuations} (c,d). The $c_f$ RMS decreases with the increase of $Re$, likely due to the decrease of the $u_{\tau}$ RMS, even though the velocities at the freestream or boundary layer edge are similar for all $Re$. As $Re$ increases, smaller vortices are generated, followed by a faster vortex advection that causes lower $c_f$ and $u_{\tau}$ RMS. $Re=10^4$ in Appendix \ref{appA} exhibits the highest contrast between the LEV area and its surroundings at the leading edge, explaining this high fluctuation across the cycle phase or higher RMS. The flow remains turbulent near the trailing edge at $Re=10^6$, mainly because the fluctuation of $u_{\tau}$ increases, or the turbulent level is stronger.

\begin{figure}[tbhp]
	\centering
        \includegraphics[width=1\textwidth]{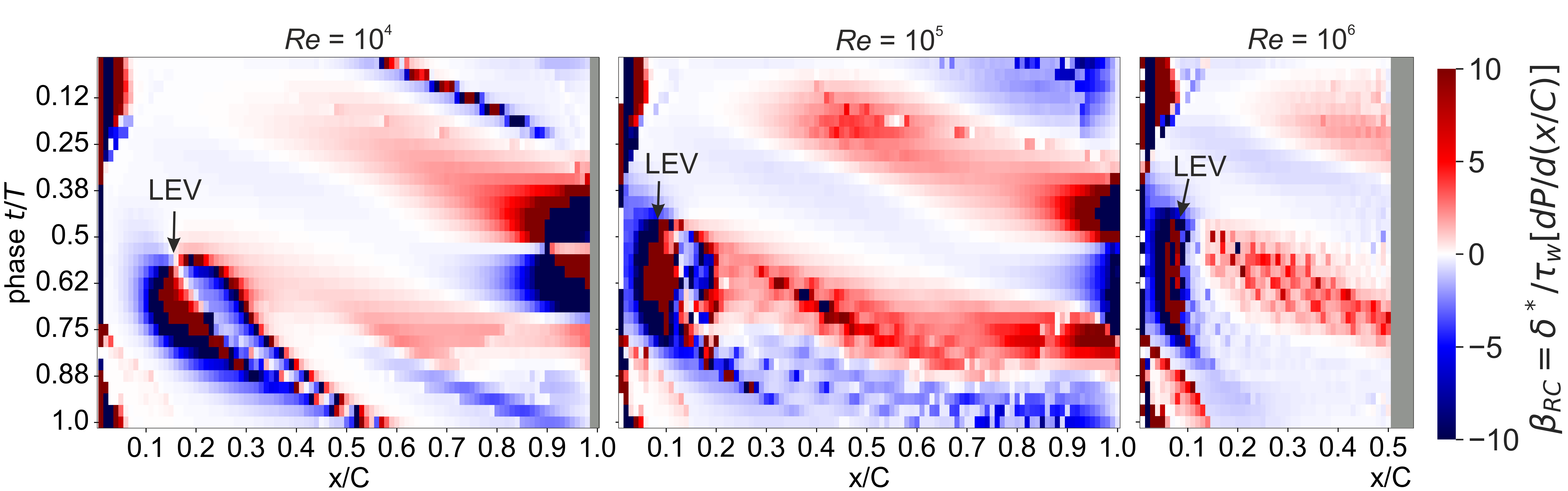}
	\caption{Clauser's pressure gradients parameter $\beta_{CL}$ normalised by displacement thickness $\delta^*$ and wall shear stress $\tau_w$ from \citet{Clauser1954}. Greys have no data, and only the first front half of chord locations are provided for $Re=10^6$ due to a high signal-to-noise ratio.}
	\label{fig:beta}
 \end{figure} 

Fig. \ref{fig:beta} illustrates the normalised pressure gradient parameter by \citet{Clauser1954} as $\beta_{RC}=\frac{\delta^*}{\tau_w}\frac{dp}{dx}$. Identical values of $\beta_{RC}$ indicate the same pressure gradient and velocity profile. Positive values indicate APG and negative values indicate FPG. The $\beta_{RC}$ are distributed similarly to the near-wall pressure gradient in Fig. \ref{fig:velo_profiles}(b) with two significant APG regions. The regions marked with "LEV" show where the LEVs are first generated, followed by the downstream trace of advection in later phases. With increasing $Re$, the LEV influence is pushed closer to the leading-edge, a sign of early LEV generation. At $Re=10^6$, the LEV region narrows significantly and is much closer to the foil's leading edge at $t/T \approx 0.5$. The LEV breakdown is also advected downstream along with the second APG region in the opposite $\beta_{RC}$ sign, as if it were disconnected from the region where the LEV was initially generated. It might explain why regions of downstream advection at $Re=10^6$ are not affected by the large LEV region nor promoted into fully 3-D turbulence. It also explains why $c_f$ and $u_{\tau}$ RMS in Fig. \ref{fig:fluctuations} are smaller with increasing $Re$. The early and narrow LEV-generated region is somehow beneficial for stabilizing the LEV breakdown at higher $Re$.

In conclusion, the FPG that promotes the expansion of the laminar-like region is not affected by the flow fluctuations caused by the breakdown of LEV. As a result, higher $Re$ may lead to more stable or undisturbed \textit{relaminarization}. It would be interesting to investigate whether this phenomenon continues to occur at even higher $Re$ in the future. 

\section{Conclusion}
We study the flow across three increasing Reynolds numbers $Re=10^4$, $10^5$, and $10^6$ at the optimum propulsive range of flapping foil $St=0.3$. We show that despite the similarity of freestream being given as forcing variables, different Reynolds number cases show unique trends for the variables inside the boundary layer. The pressure and velocity variables show similarity at the freestream, which is related to the prescribed kinematics dominant in the region far from the boundary layer. We notice that the $Cp$ trend is almost constant throughout the foil position $x/C$ but changes with phases following the heave kinematics. Meanwhile, the $u_{10C}/U_{\infty}$ variable resembles the trend of pitch kinematics where each $x/C$ varies in the vertical position $Y$ from the leading to the trailing edges per phase. Both velocity and pressure gradient at the freestream exhibit the half-cycle advection trend as the prescribed sinusoidal kinematics. 

With the increase of Reynolds numbers, the flapping foil generates more LEVs of a much smaller size. As a result, increasing Reynolds numbers show more distinct behaviours inside the boundary layer region, which are mostly influenced by the LEV creation and breakdown. During a half cycle when no vortex breakdown, the wall pressure gradient follows the same trend as the freestream pressure gradient. It means that the flapping kinematics has a strong influence on the flow. The flow is \textit{relaminarized} during the influence of strong FPG showing a growing laminar region, similarly concluded by \citet{Warnack1998}. 

Another half cycle of flapping foil flow is dominated by LEVs breakdown. This vortex breakdown dictates the downstream flow and subsequent phases. As higher $Re$ produces smaller and numerous LEVs, the advection is faster than lower $Re$. Higher $Re$ increases the chance of flow becoming turbulent and generating more separation, but on the contrary, the flow seems to be more \textit{relaminarized}. The $c_f$ RMS along the phase shows a decrease with the $Re$. It is related to the fact that the LEV-generation region at higher $Re$ becomes narrower and closer to the leading edge while the LEV advection detaches from the region where it was initially generated. This explains why \textit{relaminarization} is more stable at $Re=10^6$ as the LEV breakdown with smaller sizes does not seem to promote more 3-D turbulent downstream flow.

Variables indicating \textit{relaminarization} by FPG mentioned by \citet{Narasimha1978} seem to fail for our flapping foils studies. The reliable indications are the combination of a decrease in the cyclic fluctuation of the friction coefficient $c_f'$, an increase in $H$ while the outer flow is accelerating (positive $du_e/d(x/C)$). A more controlled study is needed to quantify the effect of these variables and the turbulent energy to accurately measure the \textit{relaminarization} criteria, which is difficult to achieve in current studies. A global scale that governs the rate of the \textit{relaminarization} region across the phases is also difficult to find. This is due to the combination of influences from the upstream, the previous phase, and the vortex breakdown (separated flow). \citet{Marusic2010} describe the challenge in scaling in the inner and outer region of the boundary layer for the case of non-equilibrium flow, which is the exact case for flapping foil due to the historical effect.

It requires numerical simulations at much higher Reynolds numbers to study the effect of fully turbulent flow on flapping foil at propulsive $St$. The fact that $c_f$ fluctuates less at increasing $Re$ signifies that it will be even harder to obtain a fully turbulent flow because the \textit{relaminarization} persists. However, we are not sure of what happens beyond $Re>10^6$.  \citet{Fukagata2023} stated that the drag reduction of 40\% due to the streamwise \textit{relaminarization} in the experiments of downstream traveling wave can still happen even at $Re_{\tau}=10^5$. Thus, there is a high possibility that even beyond $Re>10^6$, the flapping flow at propulsive $St$ will be fully dominated by 3-D turbulence. This might be of interest to drag and noise reduction studies by digging more into flapping foil flow at much higher Reynolds numbers. This \textit{relaminarization} condition somehow facilitates numerical simulations at higher $Re$ to have reduced grid sizes compared to the simulation for a stationary foil with an angle of attack at stall. 

\section*{Acknowledgement}
The Authors would like to thank the Defence Science and Technology Laboratory DSTL UK for the earlier funding and IRIDIS High-Performance Computing Facility with its associated support services at the University of Southampton. Authors would also like to thank Professor Ugo Piomelli and Dr Takafumi Nishino for informal email discussions and valuable feedback.

\section*{Declaration of interests}
The authors report no conflict of interest. 

\section*{Appendix A}\label{appA}
\begin{figure} [tbhp]
    \centering
        \includegraphics[width=1\textwidth]{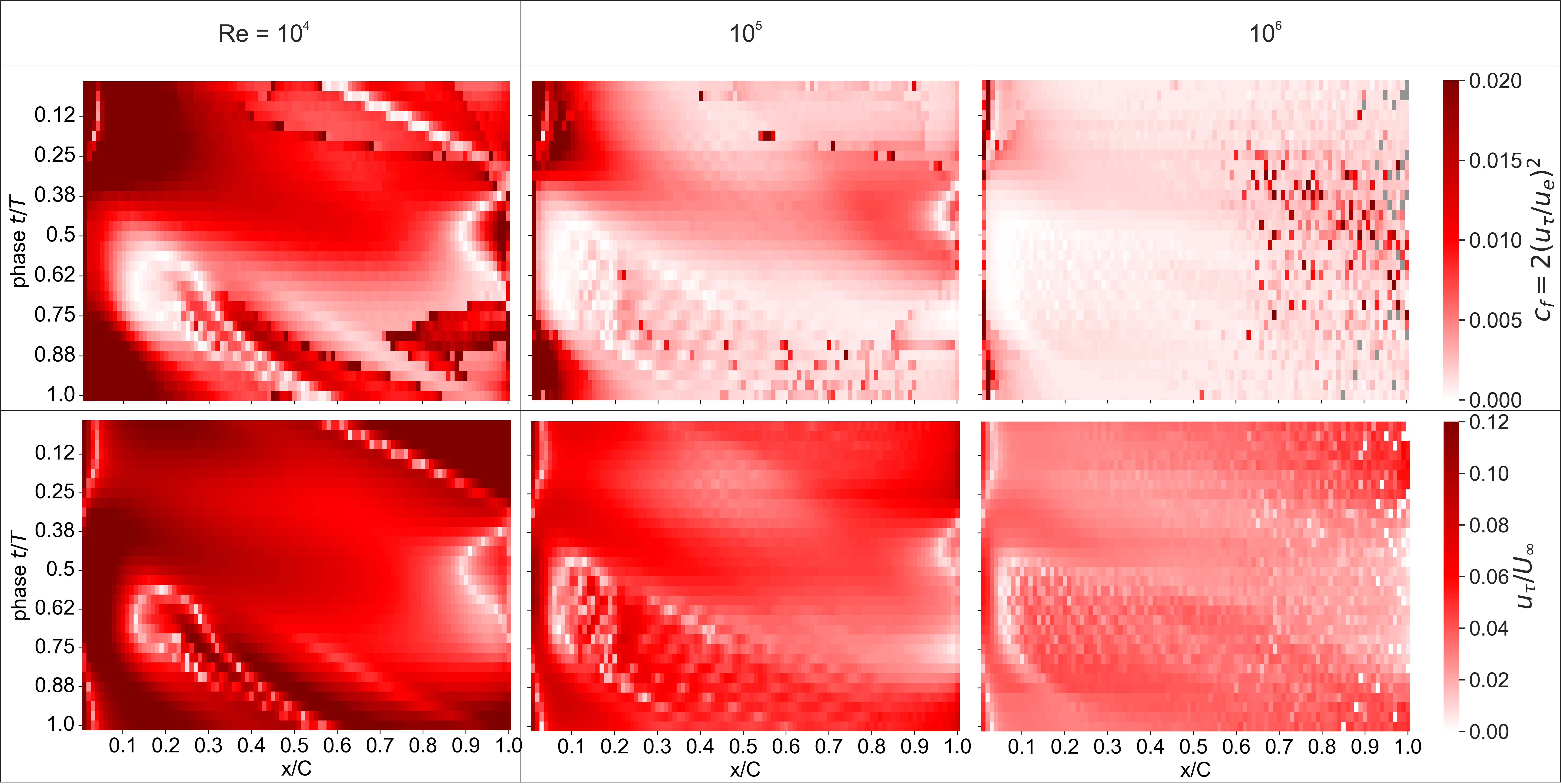} 
    \caption{Nominal values of friction coefficient $c_f$ and friction velocities $u_{\tau}$ of increasing Reynolds number presented on heatmaps along the phases $t/T$ and position on foil $x/C$. Grey cells show no values. The nominal values are decreasing with $Re$, creating a more uniform map across the phase and location, thus decreasing the fluctuation generally.}
	\label{fig:cf_utau_Re}
 \end{figure}

Nominal values of wall friction coefficient $c_f$ the friction velocity $u_{\tau}$ for different $Re$ are shown in Fig. \ref{fig:cf_utau_Re}.

\bibliography{Reference}
\end{document}